\documentclass{article}
\PassOptionsToPackage{numbers, sort}{natbib}
\usepackage[preprint]{neurips_2025}
\usepackage[utf8]{inputenc}
\usepackage[T1]{fontenc}
\usepackage{url}
\usepackage{microtype}
\usepackage{graphicx}
\usepackage{booktabs}
\usepackage{amsmath}
\usepackage{amssymb}
\usepackage{algorithm}
\usepackage{algpseudocode}
\usepackage{placeins}
\usepackage{flafter}
\usepackage{enumitem}
\usepackage[table]{xcolor}
\usepackage{multirow}
\usepackage{pifont}
\usepackage[most]{tcolorbox}
\usepackage[colorlinks=true]{hyperref}
\usepackage{todonotes}
\usepackage{fontawesome5}

\definecolor{mydeepblue}{RGB}{39, 60, 117}
\usepackage{listings}
\definecolor{promptNavy}{HTML}{1B262C}
\definecolor{promptSky}{HTML}{BBE1FA}
\newtcolorbox{promptbox}[1]{breakable, colback=promptSky!10, colframe=promptNavy, colbacktitle=promptNavy, coltitle=white, title={\scriptsize\textbf{\texttt{#1}}}, boxrule=0.6pt, left=4pt, right=4pt, top=3pt, bottom=3pt}
\newtcolorbox{findingbox}[1]{
  breakable,
  colback=blue!3,
  colframe=blue!40,
  colbacktitle=blue!10,
  coltitle=black,
  title={Finding~#1},
  fonttitle=\bfseries,
  boxrule=0.5pt,
  left=6pt, right=6pt, top=4pt, bottom=4pt
}
\hypersetup{linkcolor=mydeepblue,citecolor=mydeepblue,urlcolor=mydeepblue}

\newcommand{\method}{AHE-Beta}
\newcommand{\usr}[1]{\mbox{\method{} ($N{=}#1$)}}
\newcommand{\usrnomid}{\method{}~($N{=}8$, w/o interaction)}
\newcommand\blfootnote[1]{%
  \begingroup
  \renewcommand\thefootnote{}%
  \footnotetext{#1}%
  \endgroup
  \addtocounter{footnote}{-1}%
}

\title{Agentic Harness Engineering Beta: Requirement-Centered Evolution of Agent Harnesses}

\author{%
 \mdseries
 \textbf{Jiahang Lin\textsuperscript{1}$^{*}$},
 \textbf{Jiahao Chen\textsuperscript{1}$^{*}$},
 \textbf{Yiding Wang\textsuperscript{1}$^{*}$},
 \textbf{Shichun Liu\textsuperscript{1}},
 \textbf{Chengjun Pan\textsuperscript{2}}, \\
 \textbf{Shihan Dou\textsuperscript{1}},
 \textbf{Honglin Guo\textsuperscript{1}},
 \textbf{Yuyang Hu\textsuperscript{3}}, 
 \textbf{Changhao Jiang\textsuperscript{1}}, 
 \textbf{Zhenhua Han\textsuperscript{4}}, \\
 \textbf{Hang Yan\textsuperscript{4}}, 
 \textbf{Qi Zhang\textsuperscript{1}}, 
 \textbf{Tao Gui\textsuperscript{1}$^{\dagger}$},
 \textbf{Xuanjing Huang\textsuperscript{1}$^{\dagger}$}
 \\[4pt]
 \textsuperscript{1}Fudan University \quad
 \textsuperscript{2}Peking University \quad \\
 \textsuperscript{3}Renmin University of China \quad
 \textsuperscript{4}Shanghai Qiji Zhifeng Co., Ltd\\
 \faGithub~\href{https://github.com/Curry09/AHE-Beta.git}{\texttt{Curry09/AHE-Beta}}
}

\begin{document}
\maketitle
\blfootnote{$^{*}$Equal contributions.\ \ $^{\dagger}$Corresponding authors: \url{tgui@fudan.edu.cn}, \url{xjhuang@fudan.edu.cn}.}

\begin{abstract}

Real-world harness engineering is driven by explicit and implicit user requirements, which guides development and validation. However, existing harness evolution methods often overlook implicit requirements and interaction needs, while using a single evolution agent to modify the harness and benchmark scores to select changes. This process often produces narrow patches and leads to benchmark-specific local optima.
We introduce \textbf{Agentic Harness Engineering Beta (\method{})}, a \textbf{requirement-centered} loop that brings real-world engineering practices into harness evolution. The loop consists of three steps:
\ding{182} User-driven evaluation lets simulated users raise their requirements as issues while they work with the harness over multiple turns.
\ding{183} Team-based modification implements those requirements as generalizable changes rather than task-specific patches.\ding{184} Requirement-driven selection merges a change only when it resolves a verified requirement and passes review. Empirically, on held-out tasks, \method{} raises pass@1 from 21.9\% to 33.3\% while using 40.8\% fewer tokens than Base Harness, and achieving higher pass@1 than Claude Code, Meta-Harness, and AHE. In a blind human A/B test, real users prefer \method{} in 31 of 48 valid pairs and rate it higher on all five dimensions, matching the directional preference of the user simulator. The human A/B results show that \method{} achieves its goal of aligning harness evolution with real-world user requirements. Without further evolution, the gains transfer to GLM-5.3 and DeepSeek-V4.1-Flash with pass@1 improvements of 15.8\% and 5.3\% over Base Harness. Together, these results show that requirement-centered harness evolution is a more general method for improving agent capabilities.

\end{abstract}

\section{Introduction}

\begin{figure*}[t]
\centering
\includegraphics[width=\textwidth]{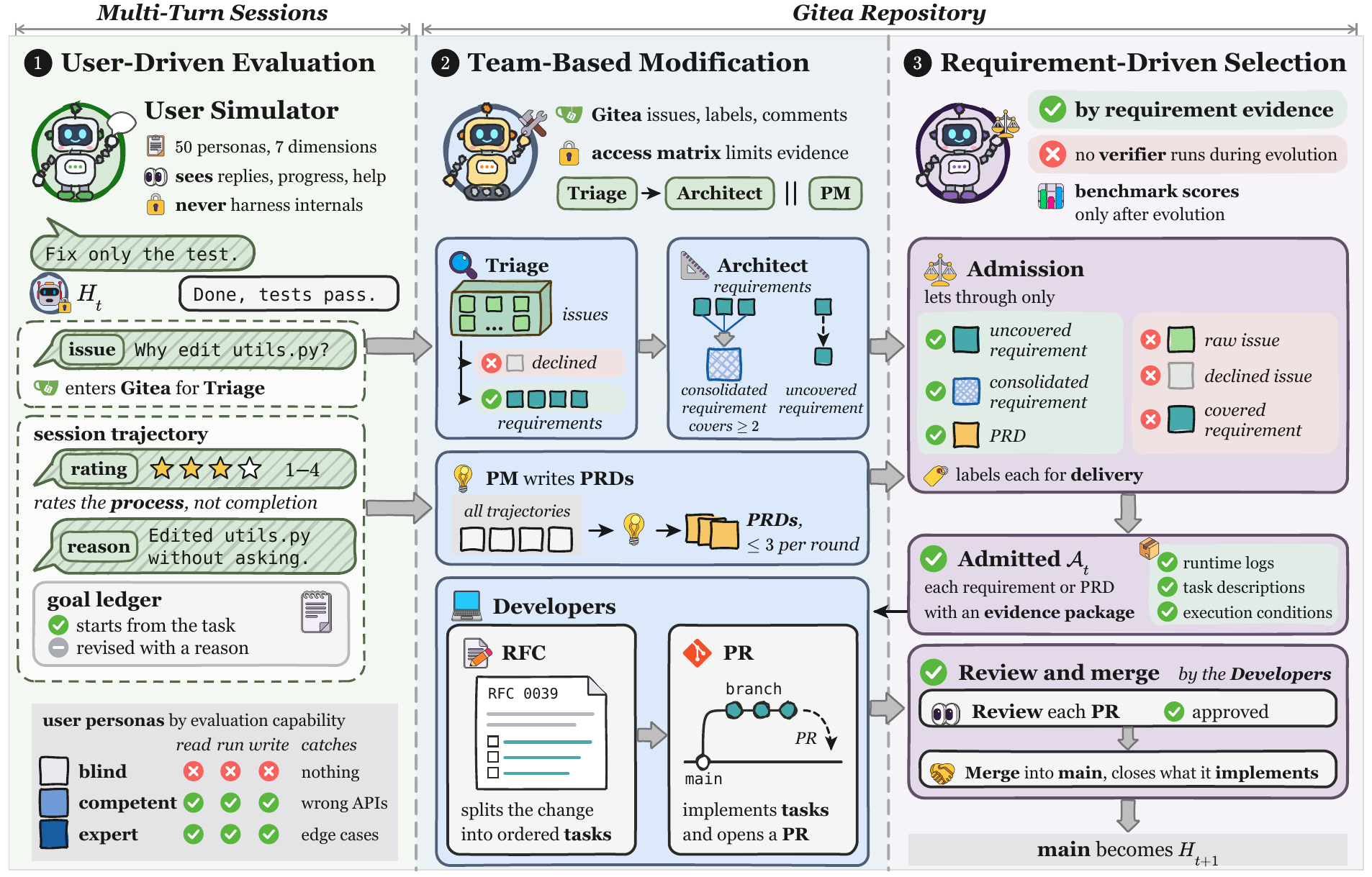}
\caption{Overview of \method{} in round $t$.
\ding{182} The user simulator works with $H_t$ on evolution tasks, files issues, and rates each session.
\ding{183} Triage, the Architect, the PM, and the Developers distill requirements from the issues, list where the harness should change, and implement them as PRs.
\ding{184} PRs with verified requirements are merged into main, which becomes $H_{t+1}$.}
\label{fig:overview}
\end{figure*}

Coding agents are increasingly deployed on long-horizon software-engineering tasks, including issue resolution in real-world repositories and multi-step terminal workflows \citep{li2025aidev,murphyhill2026claudecode,jimenez2024swebench,deng2025swebenchpro,merrill2026terminalbench}.
Performance on such tasks depends not only on the model but also on the harness, which comprises the prompts, tools, and middleware around a base model and shapes how the model executes tasks and interacts with users.
For a fixed model, behavior varies substantially across harness designs \citep{yang2024sweagent,starace2026scaffold}, and effective designs are model-specific \citep{zhang2026selfharness,lin2026updating}, this model dependence motivates harness evolution.

Real-world harness engineering is driven by user requirements, which can be explicit or implicit.
Explicit requirements are stated in user requests, while implicit requirements are unstated expectations about the results or behavior of the harness.
Engineering teams use both to guide development and validation.
Recent harness evolution methods, however, center on benchmark outcomes \citep{lin2026ahe,lee2026metaharness,zhang2026selfharness}, leaving gaps between these methods and real-world engineering practice in evaluation, modification, and selection.
In evaluation, benchmark rollouts produce execution trajectories and verifier results for predefined tasks.
These records can miss implicit user requirements, including unstated preferences about scope and cost, expectations revealed through mid-run corrections, and the need for timely responses and visible verification evidence \citep{raghavendra2026sweinteract,zhou2026swerpg}.
In modification, an evolution agent can improve benchmark scores by fixing observed failures without explicitly defining the user requirements behind those fixes \citep{lin2026ahe,lee2026metaharness,zhang2026selfharness}.
Real-world harness engineering instead judges changes by whether they resolve user problems, so higher benchmark scores alone do not establish that a change is useful.
In selection, benchmark scores alone cannot distinguish general fixes from task-specific patches.
Score-based selection can therefore reinforce overfitting and trap evolution in benchmark-specific local optima \citep{wang2026rethinking,wang2026compound}.

The key challenge in agent-driven harness evolution is to produce changes that satisfy both explicit and implicit user requirements while generalizing beyond the evolution tasks.
Our central insight is that \textit{simulated user interactions} can reveal implicit requirements that task verifiers miss and provide evidence of whether harness behavior satisfies user needs.
These requirements can guide general fixes across related failures and provide acceptance criteria independent of benchmark scores. 
We propose Agentic Harness Engineering Beta (\method{}), a requirement-centered loop that brings real-world engineering practices into each of the three steps.
\ding{182} User-driven evaluation draws feedback from multi-turn sessions in which the user simulator works with the current harness and files issues for requirements that task verifiers miss.
\ding{183} Team-based modification hands these issues to a Gitea-based team in which Triage filters out invalid and task-specific ones, the Architect merges those sharing a root cause into consolidated requirements, the Product Manager (PM) infers requirements that users leave unstated, and the Developers turn them into general fixes.
\ding{184} Requirement-driven selection ties each merge to a verified requirement and an approved review, and reserves benchmark scores for evaluation after evolution.

We evaluate \method{} on held-out tasks from Harbor-index \citep{shi2026harborindex}. Compared with Base Harness, \method{} raises pass@1 from 21.9\% to 33.3\% and user rating from 2.59 to 3.04, while using 40.8\% fewer tokens and exceeding Claude Code \citep{liu2026diveclaudecode}, Meta-Harness \citep{lee2026metaharness}, and AHE \citep{lin2026ahe} in pass@1.
In a blind human A/B test, real users prefer \method{} in 31 of 48 valid pairs and rate it higher on all five dimensions. Without further evolution, pass@1 also improves by 15.8 and 5.3 percentage points over Base Harness on GLM-5.3 and DeepSeek-V4.1-Flash, respectively.
These transfer results indicate that selecting harness modifications according to user requirements can produce improvements that generalize beyond the tasks and model used during evolution. We also find that some verifier-passing sessions never deliver a completed reply and receive the minimum user rating, showing that pass@1 alone is an incomplete measure of requirement satisfaction. Together, these results show that user requirements can guide improvements in task execution, efficiency, and real-user experience without selecting changes by benchmark score.

This paper makes two contributions:
\begin{itemize}[leftmargin=*,itemsep=2pt]
\item We propose \method{}, a harness evolution framework centered on user requirements. It combines user-driven evaluation, team-based modification, and requirement-driven selection to identify unmet needs, develop general fixes, and merge changes that resolve verified requirements.
\item We show that requirement-centered harness evolution improves pass@1, user rating, and token efficiency on held-out tasks. \method{} raises pass@1 from 21.9\% to 33.3\% over Base Harness, surpassing Claude Code, Meta-Harness, and AHE. Gains across all four domains and transfer to other models without further evolution support the generality of the evolved harness.
\end{itemize}

\section{Related Work}

\paragraph{Automated harness engineering.}
Automatic optimization has progressed from prompt and context optimization \citep{zhou2023ape,yang2023opro,pryzant2023protegi,yuksekgonul2024textgrad,fernando2023promptbreeder,khattab2023dspy,opsahlong2024mipro,agrawal2025gepa,zhang2025ace,jin2026generalistautonomousresearchhypothesistree} and learning from reflection, experience, and memory \citep{shinn2023reflexion,madaan2023selfrefine,zhao2024expel,wang2023voyager,wang2024awm,ouyang2025reasoningbank,wei2025evomemory} to agent design search and self-modifying agent code \citep{hu2024adas,zhang2024aflow,shang2024agentsquare,yin2024godelagent,zhang2025dgm,robeyns2025sica}. Recent methods evolve the complete harness and enrich the evidence for each revision with component, experience, and decision observability \citep{lin2026ahe}, prior candidates \citep{lee2026metaharness}, weakness mining \citep{zhang2026selfharness}, first-person feedback \citep{zhang2026harnesscompass}, and human demonstrations \citep{che2026demoevolve}. When task verifiers provide the feedback and benchmark scores select the changes, the evolved harnesses can overfit the evolution tasks and fail to accumulate gains \citep{wang2026rethinking,zhang2026harnesscompass,wang2026compound}. \method{} draws feedback from user interaction, merges changes by verified requirements, and reserves benchmarks for evaluation after evolution.

\paragraph{User simulation.}
Simulated users evaluate dialogue systems and agents \citep{schatzmann2007agenda,balog2025usersim}. Interactive benchmarks give them hidden goals, rules, and preferences that the agent must infer over turns, both in general domains \citep{yao2024taubench,barres2025tau2bench,qian2025userbench,wang2023mint} and in repository work \citep{king2026dialogueswe,wu2026swetogether,raghavendra2026sweinteract,peng2026icaebench,wang2026asukabench}. To recover requirements that users leave unstated, prior work studies when and what agents should ask \citep{zhang2023clarifywhen,mu2024clarifygpt,qian2024in3,wu2024humanevalcomm,vijayvargiya2025ambigswe,jin2026reqelicitgym} and keeps a persistent model of each user's goals and preferences \citep{zhou2025tomswe}. Across users, simulated feedback is aggregated into requirements documents and usability reports \citep{ataei2024elicitron,holter2026uxcascade} or used as rewards for reinforcement learning \citep{wu2025collabllm,zhou2025sweetrl,qian2025userrl,sun2025ppp}. \method{} also aggregates feedback across sessions and users, but turns unmet requirements into interpretable, revertible harness changes rather than per-user memory, documents, or model weights.

\paragraph{Multi-agent software engineering.}
Prior work assigns requirements, architecture, implementation, testing, and review to specialized roles that coordinate through intermediate artifacts and structured handoffs \citep{qian2023chatdev,hong2023metagpt,lin2025soen101,nguyen2025agilecoder}. Repository-level systems split issue resolution into localization, patching, validation, and review \citep{tao2024magis,arora2024masai}, and requirement-focused systems elicit, validate, and trace requirements from natural-language goals to code \citep{jin2024mare,jin2025iredev,chakraborty2026reqlens,chen2026tracedev}. These teams build or fix one product for a given task. \method{} applies the same division of labor to harness evolution, where Triage validates issues, the Architect consolidates requirements that share a root cause across users and sessions, the PM mines unstated requirements from all sessions, and the Developers implement them as reusable harness changes.

\section{Method}
\label{sec:method}

\method{} keeps the base model $M$ fixed and evolves only the harness $H$ around it, whose code lives in a Gitea repository. Each round lets users work with the current harness $H_t$, lets a team revise the code from their feedback, and merges the reviewed changes into $H_{t+1}$ (Figure~\ref{fig:overview}). Requirements connect the three steps. A requirement states how the harness should behave independent of any single task, with acceptance criteria and evidence from sessions.
\ding{182} User-driven evaluation (Section~\ref{sec:method:eval}) collects feedback. In each session, the user simulator uses the current harness to complete an evolution task over multiple turns, files issues along the way, and gives a user rating at the end.
\ding{183} Team-based modification (Section~\ref{sec:method:mod}) lets Triage filter the issues, the Architect and the PM distill requirements from the remaining issues and the sessions, and the Developers implement these requirements and open pull requests (PRs).
\ding{184} Requirement-driven selection (Section~\ref{sec:method:sel}) decides which changes enter the harness. A PR merges only after admission and review, so no step consults benchmark scores. Algorithm~\ref{alg:method} gives one complete run.

\begin{algorithm}[!ht]
\caption{One \method{} run.}
\label{alg:method}
\begin{algorithmic}[1]
\Require base model $M$, base harness $H_1$, personas $\mathcal{U}$, evolution tasks $\mathcal{X}$, rounds $T$
\For{$t = 1$ to $T-1$}
    \State $(\mathcal{S}_t, \mathcal{I}_t) \gets \textsc{Simulate}(M, H_t, \mathcal{U}, \mathcal{X})$ \Comment{rated sessions $\mathcal{S}_t$, issues $\mathcal{I}_t$}
    \State $\mathcal{R}_t \gets \textsc{Triage}(\mathcal{I}_t, \mathcal{S}_t, H_t)$ \Comment{requirements from accepted issues}
    \State $\mathcal{D}_t \gets \textsc{PM}(\mathcal{S}_{1:t})$ \Comment{PRDs from all sessions so far, in parallel with Triage}
    \State $(\mathcal{R}^{\mathrm{con}}_t, \mathcal{C}_t) \gets \textsc{Architect}(\mathcal{R}_t, \mathcal{S}_{1:t})$ \Comment{consolidated requirements, the requirements they cover}
    \State $\mathcal{A}_t \gets (\mathcal{R}_t \setminus \mathcal{C}_t) \cup \mathcal{R}^{\mathrm{con}}_t \cup \mathcal{D}_t$ \Comment{admitted requirements and PRDs}
    \State $H_{t+1} \gets \textsc{Developers}(H_t, \mathcal{A}_t)$ \Comment{RFC, PR, review, merge}
\EndFor
\State $\textsc{Simulate}(M, H_T, \mathcal{U}, \mathcal{X})$ \Comment{final round rates $H_T$, users only}
\State \Return $H_T$
\end{algorithmic}
\end{algorithm}

\subsection{User-Driven Evaluation}
\label{sec:method:eval}

User-driven evaluation tests the harness with a user simulator. Each session pairs an evolution task with a persona, a user profile distilled from the public writing and GitHub activity of a real developer that records traits such as the acceptance bar, the expected autonomy of the harness, and the ability to check results. The user simulator follows its persona as it works on the task with the harness. It observes the harness only through the user-facing interface, which provides replies and progress updates. Throughout the session, the persona contributes unstated preferences, the multi-turn interaction lets new goals emerge, and the issues and user rating record which requirements remain unmet.

\paragraph{Personas.}
The persona library holds 50 personas, and each describes seven dimensions such as acceptance bar, autonomy expectation, and evaluation capability, grounded in quoted evidence where the source supports it (see Appendix~\ref{app:profile}). These dimensions guide how the user interprets progress, checks results, and decides when to intervene. Only the simulator has access to the persona. The harness learns about the user's preferences through the messages they send.

\paragraph{Multi-turn interaction.}
During the interaction, the user simulator can read the harness's replies, follow its progress, and consult the available help. It cannot inspect the harness implementation. At each turn, the user simulator assesses progress and may check results using tools allowed by its persona. The user simulator first states its intended next step in its own words. It then receives the action menu, selects an option, and records any gap between its stated intent and what that option allows. This design follows Norman's account of how users form intentions, act, and evaluate system feedback \citep{norman1986cognitive}. Forming intent before selecting an action allows the user to express needs beyond the available options, while the resulting trajectory records how those needs emerge and change. The simulator keeps a goal ledger across turns, inspired by how plans guide later decisions \citep{bratman1988plans}. The ledger records the original task, new goals, and the reasons for revisions, and earlier observations and check results remain available to later judgments.

\paragraph{Feedback.}
During the interaction, the user simulator can report what went wrong or describe what it still needs from the harness. At the end of the interaction, it also provides an overall user rating with a brief explanation, based on its observations and expectations.

\subsection{Team-Based Modification}
\label{sec:method:mod}

Team-based modification turns the feedback of a round into generalizable harness changes through four roles that collaborate in the Gitea repository. Triage filters issues and keeps the requirements worth pursuing. The Architect finds requirements that share a root cause and merges them into one structural requirement. The PM finds requirements that users leave unstated across sessions and writes them as product requirements documents (PRDs). The Developers write a design document called a request for comments (RFC) for each admitted requirement or PRD and then open a PR to implement it. Triage and the Architect handle the issues that users file, and the PM handles the user ratings and summaries of sessions. The two routes run in parallel, and their output reaches the Developers only through admission (Section~\ref{sec:method:sel}).

\paragraph{Triage.}
Triage handles one issue at a time. It reads the issue, the interaction records and harness logs that the framework saves for the session that filed it, and the harness source at the commit that session used. It rejects an issue that concerns the user task rather than the harness, does not hold, or is already satisfied by the current harness. It rewrites each remaining issue as a requirement free of project names, task IDs, and specific answers, labeled with a kind, a priority, and references to session evidence. Triage does not deduplicate across issues, leaving requirements that share a root cause for the Architect to merge. When the harness logs cannot settle a judgment, for example whether the harness still runs the old request or the new one after a user interruption, Triage files a requirement that the harness record such facts instead of guessing a root cause.

\paragraph{Architect.}
After Triage finishes, the Architect reads the requirements of the round, together with the user rating and reason of every session so far, and writes consolidated requirements. Each consolidated requirement cites at least two requirements and states their relationship, common tension, structural direction, and expected leverage. It stays at the level of a mechanism and names no function or implementation step. Its \texttt{covers} field lists the requirements it absorbs, which close as merged into it, while requirements outside \texttt{covers} pass on unchanged.

\paragraph{PM.}
The PM reads the user rating, reason, and goal summary of every session so far, with open requirements and PRDs as reference only, and looks for recurring patterns, or single patterns with large impact, in how the harness serves user goals, and drills into the timeline of a session only to verify a candidate pattern. It writes at most three PRDs per round. Each PRD states the user goal, observed behaviors, experience tension, product opportunity, desired outcome, and evidence sessions, and stays at the level of a product capability. PRDs bypass the Architect and go straight to admission.

\paragraph{Developers.}
The Developers first order the admitted requirements and PRDs by the interfaces and files they share. For each requirement or PRD, the Developers write an RFC under \texttt{docs/specs/} that fixes the objective, constraints, and non-goals, compares candidate designs at different layers of the harness, and splits the selected design into two to six tasks. Each task owns a set of files disjoint from the other tasks and lists an acceptance criterion and the tasks it depends on. The Developers then implement the tasks in dependency order on a separate branch, cite the RFC and the task in code comments, run lint and tests, and open a PR.

\subsection{Requirement-Driven Selection}
\label{sec:method:sel}

Requirement-driven selection merges only changes that resolve a verified requirement and pass review. It applies two gates. The first is admission, which passes only requirements and PRDs backed by session evidence to the Developers. The second is review, which merges a PR only after its RFC passes design review and its code passes code review against the acceptance criteria of the RFC.

\paragraph{Admission.}
Admission runs once at the end of the feedback stage of each round. Admission takes the PRDs, the consolidated requirements, and the requirements outside \texttt{covers}, hands each to the Developers by adding a delivery label, and attaches an evidence package with the runtime logs of the harness for the cited sessions, the public task descriptions, and execution conditions such as budget and follow-up availability. User conversations, hidden tests, and graders stay out of the package. Raw issues, declined issues, and covered requirements never reach the Developers, so every delivered change rests on session evidence that a requirement role checked and cited.

\paragraph{Review and merge.}
Before implementation, a design review checks that the RFC preserves the objective, intervenes at the layer that owns the behavior, covers interface and state boundaries, and specifies a test that could falsify the design, and a failed check sends the RFC back for one revision. After implementation, a code review checks the PR against the acceptance criteria of the RFC, requires the full test suite to finish, and returns PASS or REWORK, and REWORK sends the PR back to the Developers. A PR merges into the main branch only when the approval is bound to the current head of the PR and the RFC still matches the reviewed design, and the merge closes the requirement or PRD that the PR implements. Every change is a git commit, so the Developers can revert any change during review and conflict resolution.

% Experiments section. Figure and table style follows arXiv:2604.25850 (AHE):
% a research-question box, RQ-named subsections, claim-sentence paragraph heads,
% captions above tables, gray group header rows, task counts in the header,
% arrow-marked metric groups, and bold best per column.
% Requires graphicx, booktabs, [table]{xcolor}, and [most]{tcolorbox}.
% Figure: figures/u32-evolution.pdf. A/B results appear in Appendix~\ref{app:ab}.

\section{Experiments}
\label{sec:experiments}

We organize our evaluation around three research questions about meeting user requirements, the contribution of user feedback and interaction beyond task verifiers, and generalization.

\begin{tcolorbox}[
    colback=blue!3,
    colframe=blue!40,
    arc=2pt,
    boxrule=0.5pt,
    left=6pt, right=6pt, top=4pt, bottom=4pt,
    title={\textbf{Research Questions}},
    fonttitle=\bfseries,
    coltitle=black,
    colbacktitle=blue!10
  ]
  \begin{enumerate}
    \setlength{\leftmargini}{1.5em}
    \setlength{\itemsep}{2pt}
    \setlength{\parsep}{0pt}
      \item \textbf{RQ1} (\S\ref{sec:experiments:main})\textbf{: Why choose requirement-centered harness evolution over verifier-driven evolution and a human-engineered harness?}
      \item \textbf{RQ2} (\S\ref{sec:experiments:divergence})\textbf{: What do user feedback and interaction contribute beyond task verifiers?}
      \item \textbf{RQ3} (\S\ref{sec:experiments:transfer})\textbf{: Do the evolved changes generalize beyond the evolution tasks and the execution model they evolved on?}
  \end{enumerate}
\end{tcolorbox}

\subsection{Setup}
\label{sec:experiments:setup}
\label{sec:setup}

\paragraph{Evolution.}
\usr{8}, \usr{16}, and \usr{32} start independently from Base Harness with the first $N$ seeds of Appendix~\ref{app:seeds} under the same six-round schedule, and we report \usr{32} as \method{}. A seed pairs an evolution task with an evolution persona (Section~\ref{sec:method:eval}). \usrnomid{} reuses the seeds of \usr{8} but runs every session without interaction, in which the user simulator reviews only the final reply and a read-only workspace patch before filing a user rating and issues. No variant receives task-verifier results or benchmark scores. The execution model and the user simulator run GLM-5.3-Flash, the harness engineering team runs GLM-5.3, and each of our evolution settings contributes one run.

\paragraph{Evaluation.}
\label{sec:experiments:metrics}
We evaluate on 38 Harbor-index \citep{shi2026harborindex} tasks spanning Coding (14), Optimization (12), Analysis (7), and Tool Use (5), listed in Appendix~\ref{app:eval}, and no evolution task instance falls inside this suite. Claude Code joins as a human-engineered harness on the same execution model. The main protocol is evaluation with interaction, in which the user simulator plays one fixed evaluation persona and replies after each complete harness reply until it ends the session or the task time limit expires. Evaluation without interaction sends the task text once and runs no user simulation. Cross-model transfer (\S\ref{sec:experiments:transfer}) evaluates the Flash-evolved harness unchanged on GLM-5.3 and on DeepSeek-V4.1-Flash (2026-10-17 version). Evaluation uses maximum reasoning effort and preserves the suite's task limits.

\paragraph{Metrics.}
\textit{Pass@1} is the audited fraction of tasks passed with one attempt per task. \textit{User rating} is the mean rating on a 1--4 scale, which the user simulator assigns as the evaluation persona without access to hidden-test results, so only evaluation with interaction produces it. \textit{Tokens} is the mean recorded execution tokens per task in millions, excluding user simulation. The A/B study examines pairwise preference separately. User rating and audit details appear in Appendix~\ref{app:details}.

\subsection{RQ1: Why choose requirement-centered harness evolution over verifier-driven evolution and a human-engineered harness?}
\label{sec:experiments:main}

\begin{findingbox}{1}
Requirement-centered harness evolution meets both explicit and implicit user requirements better than verifier-driven evolution and a human-engineered harness.
\end{findingbox}

\begin{figure}[!htb]
    \centering
    \includegraphics[width=\linewidth]{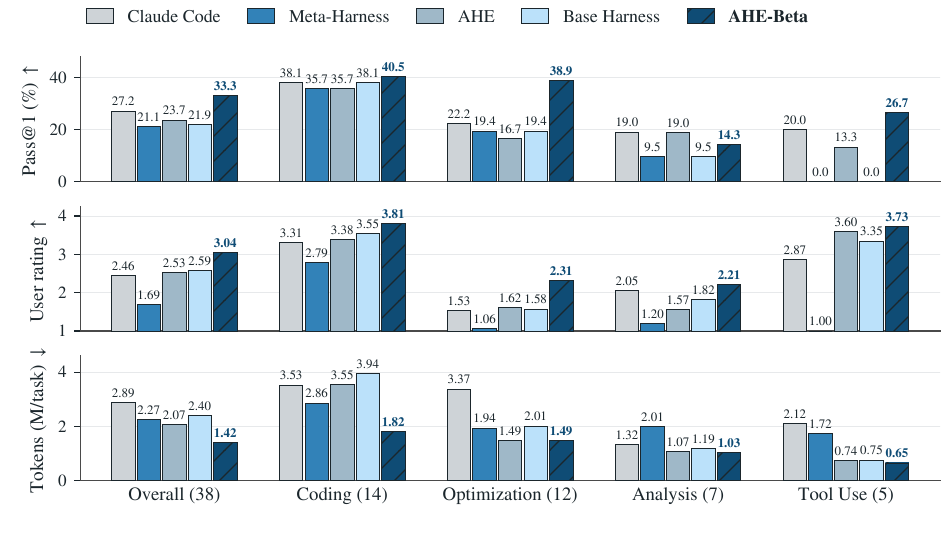}
    \caption{Main comparison on 38 held-out tasks with GLM-5.3-Flash under evaluation with interaction, by domain, with task counts in parentheses. Bars show the mean over three runs. Base Harness is the starting point of \method{}; Meta-Harness and AHE are released evolved harnesses. Hatched bars and bold labels mark \method{}. The user rating panel starts at the scale floor of 1, so a bar of zero height denotes a mean user rating of 1.00. Table~\ref{tab:main} in Appendix~\ref{app:details} lists every value.}
    \label{fig:main}
\end{figure}

\begin{figure}[!htb]
    \centering
    \includegraphics[width=\linewidth]{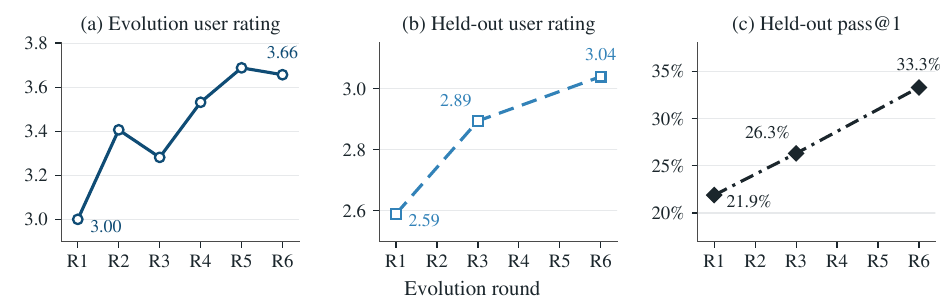}
    \caption{\method{} over recorded rounds, with R1 denoting Base Harness. Panels show (a) evolution user rating, (b) held-out user rating, and (c) held-out pass@1. Held-out observations exist only at R1, R3, and R6, and each vertical axis is truncated to its own range.}
    \label{fig:evolution}
\end{figure}

Pass@1 reflects the explicit requirements encoded in each task. \method{} raises it from 21.9\% to 33.3\% over Base Harness and exceeds Claude Code at 27.2\%, AHE at 23.7\%, and Meta-Harness at 21.1\% (Figure~\ref{fig:main}). User rating also reflects implicit requirements, and here the external harnesses diverge from \method{}. \method{} raises user rating from 2.59 to 3.04, whereas Meta-Harness, AHE, and Claude Code all fall below Base Harness at 1.69, 2.53, and 2.46. The gains of \method{} also hold per task, with higher pass@1 than Base Harness on 12 tasks and lower on 3, and higher mean user rating on 16 tasks and lower on none. Across evolution rounds, held-out user rating rises from 2.59 at R1 to 3.04 at R6, and pass@1 rises from 21.9\% to 33.3\% (Figure~\ref{fig:evolution}). On tasks it still fails, \method{} often reaches later stages than Base Harness (Appendix~\ref{app:unmet-requirements}). Requirement-centered evolution thus meets explicit and implicit requirements together and accumulates these gains across rounds, which neither verifier-driven evolution nor human engineering achieves.\label{sec:evolution-dynamics}

\begin{findingbox}{2}
Real users also prefer \method{}, in the same direction as the user simulator, showing that the evolved harness improves usability for real users.
\end{findingbox}

A blind A/B study tests whether these gains hold for real users. On six coding tasks, real users prefer \method{} in 31 of 48 valid pairs and score it higher on all five dimensions (Table~\ref{tab:ab-main}), so the evolved harness improves usability for real users. This preference matches the direction of the user simulator, whose user rating rises from 2.59 to 3.04 on held-out tasks, which supports using the user simulator to drive evolution and evaluation. The largest gain is in process understandability, which rises from 4.13 to 5.33, consistent with the progress changes traced in Appendix~\ref{app:cases}.

\begin{table}[!htb]
    \centering
    \small
    \setlength{\tabcolsep}{3pt}
    \renewcommand{\arraystretch}{1.05}
    \setlength{\abovecaptionskip}{2pt}
    \setlength{\belowcaptionskip}{2pt}
    \caption{Blind human A/B study on 48 valid pairs. Mean scores use a 1--7 scale, with higher scores indicating better experience. Overall averages the five displayed dimension means, while its Higher, Equal, and Lower counts report overall preference, including nine ties and one neither response under Equal. Other counts compare the paired dimension scores. Appendix~\ref{app:ab} defines the five dimensions, and usefulness and control each lack one rating.}
    \label{tab:ab-main}
    \begin{tabular}{lrr*{3}{>{\centering\arraybackslash}p{1.3cm}}}
        \toprule
        & \multicolumn{2}{c}{Mean score} & \multicolumn{3}{c}{\method{} against Base Harness} \\
        \cmidrule(lr){2-3}\cmidrule(lr){4-6}
        Measure & Base Harness & \textbf{\method{}} & Higher & Equal & Lower \\
        \midrule
        Overall & 5.39 & \textbf{6.01} & 31 & 10 & 7 \\
        \midrule
        Process understandability & 4.13 & \textbf{5.33} & 29 & 15 & 4 \\
        Usefulness & 5.40 & \textbf{6.15} & 16 & 26 & 5 \\
        Control & 5.23 & \textbf{5.77} & 18 & 25 & 4 \\
        User burden & 5.98 & \textbf{6.27} & 17 & 25 & 6 \\
        Ease of use & 6.21 & \textbf{6.54} & 14 & 29 & 5 \\
        \bottomrule
    \end{tabular}
\end{table}

\FloatBarrier
\subsection{RQ2: What do user feedback and interaction contribute beyond task verifiers?}
\label{sec:experiments:divergence}

\begin{findingbox}{3}
User feedback exposes implicit requirements and acts as a supervisor that raises pass@1.
\end{findingbox}

\label{sec:interaction}

\begin{table}[!htb]
    \centering
    \small
    \setlength{\tabcolsep}{4pt}
    \renewcommand{\arraystretch}{1.05}
    \setlength{\abovecaptionskip}{2pt}
    \setlength{\belowcaptionskip}{2pt}
    \caption{Evolution with and without interaction, evaluated on the 38 held-out Flash tasks. \usr{8} runs six of its eight evolution tasks with interaction, and \usrnomid{} runs none. All rows use evaluation with interaction, and units and bold follow Table~\ref{tab:main}.}
    \label{tab:interaction}
    \begin{tabular}{lrr}
        \toprule
        Setting & Pass@1\,$\uparrow$ & Tokens\,$\downarrow$ \\
        \midrule
        Base Harness & 21.9\% & 2.40 \\
        \usrnomid{} & 23.7\% & 3.15 \\
        \usr{8} & \textbf{28.9\%} & \textbf{1.71} \\
        \bottomrule
    \end{tabular}
\end{table}

During evolution, interaction exposes requirements that a single final review misses. \usr{8} reaches 28.9\% pass@1 against 23.7\% for \usrnomid{} with 45.7\% fewer tokens (Table~\ref{tab:interaction}). The savings also hold on tasks that both variants pass, since \usr{8} makes fewer model calls with shorter inputs and its stage checkpoints make progress and remaining work explicit. Interaction thus lets the user simulator raise mid-run corrections and requirements that emerge during the session, and the harness evolved on them reaches higher pass@1 with fewer tokens.

During evaluation, the user simulator acts as a supervisor that checks each reply and pushes the harness to continue or correct its work. Without it, every harness loses pass@1, yet \method{} still achieves the highest pass@1 at the lowest token usage (Table~\ref{tab:noniter}), so its advantage does not stem from user assistance during evaluation. Final ratings also expose reply delivery, a requirement that verifiers do not check. Of the 137 passed sessions, 34 deliver no completed reply and receive the minimum user rating of 1, because the verifier reads the final workspace while the user simulator sees only what the harness reports. Six of them come from \usr{16}, including Adapt 1, so its user rating stays below Base Harness despite its higher pass@1 (Figure~\ref{fig:coverage}). Verifier feedback therefore misses implicit requirements such as whether results reach the user, which motivates evolving \method{} on user feedback.

\begin{table}[!htb]
    \centering
    \small
    \setlength{\tabcolsep}{6pt}
    \renewcommand{\arraystretch}{1.05}
    \setlength{\abovecaptionskip}{2pt}
    \setlength{\belowcaptionskip}{2pt}
    \caption{User simulation during evaluation on 38 held-out tasks with GLM-5.3-Flash. Evaluation without interaction sends the task once and provides no simulated-user follow-up, so it produces no user rating. Tokens are execution tokens in millions per task, excluding user simulation. Bold marks the best pass@1 and lowest token usage within each protocol.}
    \label{tab:noniter}
    \begin{tabular}{llrr}
        \toprule
        Harness & Interaction & Pass@1\,$\uparrow$ & Tokens\,$\downarrow$ \\
        \midrule
        Claude Code & with & 27.2\% & 2.89 \\
        Meta-Harness & with & 21.1\% & 2.26 \\
        \textbf{\method{}} & with & \textbf{33.3\%} & \textbf{1.42} \\
        \midrule
        Claude Code & w/o & 23.7\% & 3.24 \\
        Meta-Harness & w/o & 13.2\% & 2.53 \\
        \textbf{\method{}} & w/o & \textbf{26.3\%} & \textbf{1.55} \\
        \bottomrule
    \end{tabular}
\end{table}

\FloatBarrier
\subsection{RQ3: Do the evolved changes generalize beyond the evolution tasks and the execution model they evolved on?}
\label{sec:experiments:transfer}

\begin{findingbox}{4}
Requirement-driven selection yields generalizable improvements across domains and execution models.
\end{findingbox}

We test generalization along two axes, held-out task domains and execution models unseen during evolution. \method{} raises both pass@1 and user rating over Base Harness in all four domains while using fewer execution tokens in each (Figure~\ref{fig:main}). The largest gains come in Tool Use, where pass@1 rises from 0.0\% to 26.7\%, and in Optimization, where it rises from 19.4\% to 38.9\%. The evolved changes therefore generalize beyond the evolution tasks.

Applied unchanged to GLM-5.3 and DeepSeek-V4.1-Flash, \method{} raises pass@1 by 15.8 and 5.3 points and user rating from 1.70 to 3.03 and from 3.33 to 3.39 (Table~\ref{tab:transfer}). The evolved changes therefore do not depend on the execution model they evolved on.

\begin{table}[!htb]
    \centering
    \small
    \setlength{\tabcolsep}{4pt}
    \renewcommand{\arraystretch}{1.05}
    \setlength{\abovecaptionskip}{2pt}
    \setlength{\belowcaptionskip}{2pt}
    \caption{Cross-model transfer of the Flash-evolved harness, paired against Base Harness on the same execution model. GLM-5.3-Flash is the evolution model, and the other two bases receive the harness without in-domain re-evolution. Pass@1 and tokens use all 38 tasks, and user rating uses paired rated tasks (38, 37, and 36 from top to bottom). Bold marks the better value in each pair.}
    \label{tab:transfer}
    \begin{tabular}{l @{\hspace{8pt}} rr @{\hspace{8pt}} rr @{\hspace{8pt}} rr}
        \toprule
        & \multicolumn{2}{c}{Pass@1\,$\uparrow$} & \multicolumn{2}{c}{User rating\,$\uparrow$} & \multicolumn{2}{c}{Tokens\,$\downarrow$} \\
        \cmidrule(lr){2-3}\cmidrule(lr){4-5}\cmidrule(lr){6-7}
        Execution model & Base & \textbf{\method{}} & Base & \textbf{\method{}} & Base & \textbf{\method{}} \\
        \midrule
        GLM-5.3-Flash & 21.9\% & \textbf{33.3\%} & 2.59 & \textbf{3.04} & 2.40 & \textbf{1.42} \\
        GLM-5.3 & 23.7\% & \textbf{39.5\%} & 1.70 & \textbf{3.03} & \textbf{1.91} & 1.99 \\
        DeepSeek-V4.1-Flash & 36.8\% & \textbf{42.1\%} & 3.33 & \textbf{3.39} & 3.80 & \textbf{2.81} \\
        \bottomrule
    \end{tabular}
\end{table}

\FloatBarrier

\section{Conclusion}

We introduced \method{}, which brings user interaction into agent-driven harness evolution. A user simulator exposes problems through multi-turn use, Triage verifies them, an Architect consolidates them by root cause, a PM discovers unstated requirements, and Developers implement changes whose effects are tested through later use. This Gitea-based loop connects user experience to continued harness-level evolution while keeping the base model fixed. Experiments show that on 38 held-out tasks, \method{} raises pass@1 from 21.9\% to 33.3\% and user rating from 2.59 to 3.04 over Base Harness with 40.8\% fewer tokens, above Meta-Harness, Claude Code, and AHE. By extending automation to user participation and requirement discovery, \method{} takes a step toward automating real-world harness engineering. Future work includes feedback from real deployments, personalized harness branches, and joint harness and model evolution.

\section*{Limitations}

Limited by time and computational resources, we study evolution with up to 32 scenarios under a six-round schedule, using one simulator model and a selected set of personas complemented by a human A/B study. Broader scenario coverage, more simulator models and persona compositions, and continued iteration could uncover additional requirements, and their effects on harness quality and computational cost remain to be evaluated at larger scales.

\method{} steers harness evolution toward user needs and may narrow the gap between evaluation and practice. Biased profiles may favor some groups, while an overly proactive harness may act beyond what users allow when requirements are unclear \citep{ji2026underspecbench}.

\bibliographystyle{unsrtnat}
\bibliography{references}

\appendix

\section{Personas}
\label{app:profile}

\begin{itemize}[leftmargin=*,itemsep=2pt]
\item \textbf{Sources.} The library holds 50 cards. 32 are extracted from first-person writing, including blog posts, tweet threads, interview transcripts, conference talks, and newsletters. 18 are extracted from GitHub repositories, issues, and PRs in the domains of SWE-bench and Terminal-Bench repositories.
\item \textbf{Dimensions.} Each card sets seven dimensions, and interaction style among them comprises delegation, supervision, and response to wrong results. The user simulator receives each value as a fixed sentence in its prompt, and Table~\ref{tab:dimensions} defines every dimension and value. Evaluation capability and domain are grounded in all 50 cards. Acceptance bar, delegation, supervision, trust, and autonomy expectation are grounded in the 32 first-person cards. For the 18 GitHub cards, acceptance bar and trust are derived from evaluation capability and domain depth, and the other ungrounded values are sampled. Response to wrong results is grounded in 45 cards and time temperament in 6.
\item \textbf{Evolution personas.} The 32 seeds draw on 15 cards, 10 first-person and 5 GitHub, and all 15 have competent or expert evaluation capability (Table~\ref{tab:personas}). Card gh-sci-03, which plays the pydicom seed of Figure~\ref{fig:interaction-case}, comes from one xarray performance issue that gives a minimal reproduction and a root-cause trace and leaves the fix to the maintainers, which grounds its evaluation capability, domain, and response to wrong results.
\item \textbf{Evaluation persona.} Held-out evaluation uses one fixed persona outside the library (last row of Table~\ref{tab:personas}). It sees completed replies and a read-only diff of the workspace and cannot run commands.
\end{itemize}

\begin{table}[ht]
    \centering
    \footnotesize
    \setlength{\tabcolsep}{4pt}
    \renewcommand{\arraystretch}{1.15}
    \caption{Persona dimensions, one row per column of Table~\ref{tab:personas}. Value meanings paraphrase the fixed sentence that the user simulator receives for each value, and Appendix~\ref{app:prompts:user} gives these sentences verbatim.}
    \label{tab:dimensions}
    \begin{tabular}{@{}l p{0.25\linewidth} p{0.58\linewidth}@{}}
        \toprule
        Column & Meaning & Values \\
        \midrule
        Evaluation & Evaluation capability: how well the user checks the output of the harness. & \textbf{blind}: skims the output and trusts it. \textbf{competent}: runs tests, reads diffs, and spots plausible but wrong APIs. \textbf{expert}: verifies systematically and catches missing edge cases and boundary bugs. \\
        Acceptance & Acceptance bar: the lowest quality the user accepts. & \textbf{runs}: the result runs without error. \textbf{functional}: it works on the normal case the user tried. \textbf{idiomatic}: it follows modern idiomatic practice. \textbf{rigorous}: it handles types and boundary cases. \\
        Delegation & Interaction style: how much context the user writes into a request. & \textbf{one-liner}: one line with little context. \textbf{half context}: background but no acceptance criteria. \textbf{spec level}: a full specification with the information the harness needs. \\
        Supervision & Interaction style: how much of the ongoing work the user needs to see. & \textbf{every step}: grows uneasy during long opaque stretches and asks for visibility. \textbf{phase check}: wants a signal at key moments that the work is on track. \textbf{result only}: cares about the final result alone. \\
        Response & Interaction style: what the user does when a result is wrong. & \textbf{feedback}: describes the problem and asks for a fix. \textbf{handover}: takes over the remaining work. \textbf{resend}: sends the request again, possibly rephrased. \textbf{give up}: abandons the task quickly. \textbf{file issue}: stops and files an issue after repeated failures. \\
        Trust & Trust: how far the user believes claims of the harness without proof. & \textbf{blind}: trusts by default. \textbf{balanced}: starts neutral and grows skeptical after repeated failures. \textbf{skeptical}: doubts by default and expects proof. \\
        Domain & Domain and stack: the technical area, tools, depth, and weak spots of the user. & Area, tools, and weak spots are free text, and depth is novice, proficient, or senior. Table~\ref{tab:personas} shows the area only. \\
        Autonomy & Autonomy expectation: how much the user lets the harness decide and act alone. & \textbf{advisory}: wants suggestions and keeps control. \textbf{collaborative}: works with the harness rather than commanding it. \textbf{agentic}: lets the harness drive and act on its own. \\
        Time & Time temperament: how long the user tolerates waiting and silence. & \textbf{patient}: a long fuse for slow or quiet stretches. \textbf{steady}: ordinary patience that wears down over repeated idle stretches. \textbf{impatient}: a short fuse that pushes the user to change course sooner. \\
        \bottomrule
    \end{tabular}
\end{table}

\begin{table}[ht]
    \centering
    \scriptsize
    \setlength{\tabcolsep}{3pt}
    \caption{Personas used in evolution and evaluation, with columns defined in Table~\ref{tab:dimensions}. sr marks first-person cards and gh marks GitHub cards. Italic values are sampled or derived, and all other card values are grounded in the source. The last row is the hand-set evaluation persona.}
    \label{tab:personas}
    \begin{tabular}{llllllllll}
        \toprule
        Card & Evaluation & Acceptance & Delegation & Supervision & Response & Trust & Domain & Autonomy & Time \\
        \midrule
        sr-01 & expert & idiomatic & spec level & phase check & handover & skeptical & dev tooling & collaborative & \textit{patient} \\
        sr-02 & expert & idiomatic & spec level & phase check & feedback & balanced & dev tooling & collaborative & \textit{steady} \\
        sr-08 & expert & idiomatic & spec level & phase check & feedback & balanced & dev tooling & collaborative & \textit{steady} \\
        sr-15 & competent & functional & spec level & phase check & feedback & balanced & dev tooling & collaborative & \textit{steady} \\
        sr-16 & competent & runs & half context & phase check & resend & balanced & web frontend & collaborative & \textit{impatient} \\
        sr-19 & competent & functional & half context & phase check & feedback & balanced & infra and DevOps & collaborative & \textit{steady} \\
        sr-20 & competent & functional & half context & phase check & feedback & balanced & dev tooling & collaborative & \textit{patient} \\
        sr-26 & expert & idiomatic & spec level & result only & feedback & balanced & dev tooling & agentic & patient \\
        sr-27 & expert & idiomatic & spec level & phase check & feedback & balanced & systems runtime & collaborative & patient \\
        sr-30 & expert & idiomatic & spec level & phase check & feedback & balanced & dev tooling & collaborative & \textit{steady} \\
        gh-db-01 & competent & \textit{idiomatic} & \textit{spec level} & \textit{every step} & \textit{feedback} & \textit{balanced} & databases & \textit{collaborative} & \textit{steady} \\
        gh-de-03 & competent & \textit{idiomatic} & \textit{spec level} & \textit{result only} & \textit{file issue} & \textit{balanced} & data engineering & \textit{collaborative} & \textit{patient} \\
        gh-ml-03 & competent & \textit{idiomatic} & \textit{half context} & \textit{phase check} & \textit{file issue} & \textit{balanced} & ML and AI & \textit{advisory} & \textit{steady} \\
        gh-sci-03 & competent & \textit{idiomatic} & \textit{one-liner} & \textit{result only} & file issue & \textit{balanced} & scientific computing & \textit{agentic} & \textit{impatient} \\
        gh-viz-03 & competent & \textit{idiomatic} & \textit{one-liner} & \textit{result only} & file issue & \textit{balanced} & data visualization & \textit{agentic} & \textit{patient} \\
        \midrule
        eval & competent & functional & spec level & phase check & feedback & balanced & Python & collaborative & steady \\
        \bottomrule
    \end{tabular}
\end{table}

\section{Seeds}
\label{app:seeds}

\begin{itemize}[leftmargin=*,itemsep=2pt]
\item \textbf{Composition.} A seed pairs an evolution task with a persona and runs with or without interaction (Table~\ref{tab:seeds}). The 32 tasks comprise SWE-bench (9) \citep{jimenez2024swebench}, Terminal-Bench (8) \citep{merrill2026terminalbench}, mechanism tasks (8), NL2Repo (4) \citep{ding2025nl2repo}, and TUA-Bench (3). Each mechanism task exercises one harness mechanism through ordinary work, namely multi-agent review, periodic monitoring, goal discovery, direction negotiation, approval gating, constraint memory, replanning, and revision from checks. The first 8 seeds contain 2, 2, 2, 1, and 1 and the first 16 contain 4, 4, 4, 2, and 2 from these sources in the same order. 25 seeds run with interaction and 7 without interaction.
\item \textbf{Schedule.} Each of the six rounds runs every seed of a variant once, and the seed order rotates by one position per round.
\item \textbf{Isolation.} Evaluation excludes seed instances and their repositories.
\end{itemize}

\begin{table}[ht]
    \centering
    \scriptsize
    \setlength{\tabcolsep}{4pt}
    \caption{The 32 seeds in run order. Rules after seeds 8 and 16 close the seed sets of \usr{8} and \usr{16}, and \usrnomid{} reuses seeds 1 to 8. Interaction marks whether a seed runs with interaction or w/o interaction (Section~\ref{sec:method:eval}), and persona IDs refer to Table~\ref{tab:personas}.}
    \label{tab:seeds}
    \begin{tabular}{rllll}
        \toprule
        \# & Source & Task & Interaction & Persona \\
        \midrule
        1 & SWE-bench & \texttt{pvlib\_\_pvlib-python-1478} & with & sr-15 \\
        2 & Terminal-Bench & \texttt{production-planning} & w/o & sr-15 \\
        3 & TUA-Bench & \texttt{010-pivot-product-revenue} & with & sr-20 \\
        4 & mechanism & \texttt{superset-goal-discovery} & with & sr-27 \\
        5 & SWE-bench & \texttt{pylint-dev\_\_astroid-2240} & with & sr-15 \\
        6 & Terminal-Bench & \texttt{kv-live-surgery} & with & sr-19 \\
        7 & NL2Repo & \texttt{tinydb} & w/o & sr-15 \\
        8 & mechanism & \texttt{sqlfluff-subagent-review} & with & sr-02 \\
        \midrule
        9 & SWE-bench & \texttt{marshmallow-code\_\_marshmallow-1229} & with & sr-02 \\
        10 & SWE-bench & \texttt{sqlfluff\_\_sqlfluff-2336} & w/o & sr-15 \\
        11 & Terminal-Bench & \texttt{vf2-speedup-networkx} & with & sr-27 \\
        12 & Terminal-Bench & \texttt{pretrain-shard-corruption} & with & gh-ml-03 \\
        13 & TUA-Bench & \texttt{091-add-receipts-bookkeeping} & with & sr-30 \\
        14 & NL2Repo & \texttt{flask-restful} & w/o & sr-15 \\
        15 & mechanism & \texttt{permission-schema-migration} & with & sr-01 \\
        16 & mechanism & \texttt{plan-blocked-dependency} & with & sr-27 \\
        \midrule
        17 & SWE-bench & \texttt{pydicom\_\_pydicom-1334} & with & gh-sci-03 \\
        18 & SWE-bench & \texttt{pyvista\_\_pyvista-3750} & with & gh-viz-03 \\
        19 & SWE-bench & \texttt{astroid-1030} & with & sr-08 \\
        20 & SWE-bench & \texttt{sqlfluff-3608} & with & gh-de-03 \\
        21 & SWE-bench & \texttt{marshmallow-2123} & with & sr-02 \\
        22 & Terminal-Bench & \texttt{payments-pipeline-fix} & w/o & sr-15 \\
        23 & Terminal-Bench & \texttt{wal-recovery-ordering} & with & gh-de-03 \\
        24 & Terminal-Bench & \texttt{vba-userform-port} & with & sr-19 \\
        25 & Terminal-Bench & \texttt{react-lead-form} & with & sr-16 \\
        26 & TUA-Bench & \texttt{080-convert-novel-epub} & with & sr-20 \\
        27 & NL2Repo & \texttt{cookiecutter} & w/o & sr-15 \\
        28 & NL2Repo & \texttt{tqdm} & w/o & sr-15 \\
        29 & mechanism & \texttt{superset-ci-loop} & with & sr-26 \\
        30 & mechanism & \texttt{dbt-direction-negotiation} & with & sr-08 \\
        31 & mechanism & \texttt{memory-project-constraints} & with & gh-db-01 \\
        32 & mechanism & \texttt{review-verifier-catches-regression} & with & sr-30 \\
        \bottomrule
    \end{tabular}
\end{table}

\section{Evaluation Tasks}
\label{app:eval}

Revision 1.4 of Harbor-index \citep{shi2026harborindex} repackages 80 tasks from other benchmarks into one format, and each task keeps its tests, reference answers, and judges in a separate verifier environment that the agent cannot access. We selected 43 tasks from two categories in Harbor-Index v1.4: \emph{Software Engineering} and \emph{Agents, Tools \& Systems}. The main comparisons use a fixed subset of 38 tasks, omitting five whose environments were difficult to configure reliably in our evaluation setup. We group the retained tasks by their primary objectives into Coding, Optimization, Analysis, and Tool Use (Table~\ref{tab:eval-tasks}). They come from SWE-bench Verified \citep{jimenez2024swebench}, SWE-bench Pro \citep{deng2025swebenchpro}, SWE-smith \citep{yang2025swesmith}, SWT-Bench Verified \citep{mundler2024swtbench}, FeatureBench \citep{zhou2026featurebench}, BigCodeBench \citep{zhuo2024bigcodebench}, Terminal-Bench \citep{merrill2026terminalbench}, USACO \citep{shi2024usaco}, AlgoTune \citep{press2025algotune}, GSO \citep{shetty2025gso}, GAIA \citep{mialon2023gaia}, SkillsBench \citep{li2026skillsbench}, WideSearch \citep{wong2025widesearch}, and Gaia2 \citep{froger2026gaia2}.

\begin{itemize}[leftmargin=*,itemsep=2pt]
\item \textbf{Coding (14).} The agent fixes bugs, adds features, writes tests, or builds programs from a specification, and hidden tests decide the outcome.
\item \textbf{Optimization (12).} The agent speeds up existing code without changing its output, and a pass requires a runtime threshold set against a reference solution.
\item \textbf{Analysis (7).} The agent finds or computes an answer from the web, documents, or data and delivers it as text or a file such as a spreadsheet, which the verifier checks against a reference.
\item \textbf{Tool Use (5).} The agent operates simulated apps such as email, calendar, and shopping through tool calls, and the verifier compares its write actions with reference actions. Adapt tasks change the situation during the run, Timed tasks require actions at set times, and the Ambiguous task asks the agent to check with the user when unsure.
\end{itemize}

\begin{table}[ht]
    \centering
    \scriptsize
    \setlength{\tabcolsep}{4pt}
    \caption{The 38 held-out Harbor-index tasks by domain, with task counts in parentheses.}
    \label{tab:eval-tasks}
    \begin{tabular}{ll}
        \toprule
        Task & Source \\
        \midrule
        \rowcolor{gray!20} \multicolumn{2}{l}{\textbf{Coding (14)}} \\
        \texttt{build-word2vec-pipeline} & BigCodeBench \\
        \texttt{featurebench-add-feature-mlflow-unity-catalog} & FeatureBench \\
        \texttt{featurebench-add-feature-xarray-backend-chunks} & FeatureBench \\
        \texttt{swebenchpro-fix-file-suffix-chooser} & SWE-bench Pro \\
        \texttt{swebenchpro-improve-process-signal-messages} & SWE-bench Pro \\
        \texttt{swebenchverified-fix-annotate-xy-array-copy} & SWE-bench Verified \\
        \texttt{swebenchverified-fix-django-mti-parent-link} & SWE-bench Verified \\
        \texttt{swebenchverified-fix-django-union-queryset-order} & SWE-bench Verified \\
        \texttt{swebenchverified-fix-span-selector-axes-limits} & SWE-bench Verified \\
        \texttt{swebenchverified-fix-sphinx-literal-nitpick} & SWE-bench Verified \\
        \texttt{swesmith-fix-oauth1-header-params} & SWE-smith \\
        \texttt{swtbenchverified-test-runserver-zero-address} & SWT-Bench Verified \\
        \texttt{tb-make-doom-for-mips} & Terminal-Bench \\
        \texttt{usaco-assign-cows-to-barns} & USACO \\
        \rowcolor{gray!20} \multicolumn{2}{l}{\textbf{Optimization (12)}} \\
        \texttt{algotune-optimize-lti-sim} & AlgoTune \\
        \texttt{algotune-optimize-matrix-sqrt} & AlgoTune \\
        \texttt{algotune-optimize-ode-seirs} & AlgoTune \\
        \texttt{algotune-optimize-outer-product} & AlgoTune \\
        \texttt{algotune-simplex-projection-speedup} & AlgoTune \\
        \texttt{gso-speedup-hf-datasets} & GSO \\
        \texttt{gso-speedup-numpy-strings} & GSO \\
        \texttt{gso-speedup-pandas-merge} & GSO \\
        \texttt{gso-speedup-pandas-period-fmt} & GSO \\
        \texttt{gso-speedup-pandas-seq-to-range} & GSO \\
        \texttt{gso-speedup-pillow-filter} & GSO \\
        \texttt{gso-speedup-pydantic-enum} & GSO \\
        \rowcolor{gray!20} \multicolumn{2}{l}{\textbf{Analysis (7)}} \\
        \texttt{gaia-compare-sciencedirect-domains} & GAIA \\
        \texttt{gaia-find-chess-winning-move} & GAIA \\
        \texttt{gaia-find-flavor-graveyard-rhyme} & GAIA \\
        \texttt{skillsbench-model-investment-shock-gdp} & SkillsBench \\
        \texttt{skillsbench-ocr-receipts-to-excel} & SkillsBench \\
        \texttt{tb-dna-insert} & Terminal-Bench \\
        \texttt{widesearch-list-bri-projects-2025} & WideSearch \\
        \rowcolor{gray!20} \multicolumn{2}{l}{\textbf{Tool Use (5)}} \\
        \texttt{gaia2-adapt-hard-1} & Gaia2 \\
        \texttt{gaia2-adapt-hard-2} & Gaia2 \\
        \texttt{gaia2-ambiguous} & Gaia2 \\
        \texttt{gaia2-timed-1} & Gaia2 \\
        \texttt{gaia2-timed-2} & Gaia2 \\
        \bottomrule
    \end{tabular}
\end{table}

\section{Role Prompts and Repository Protocol}
\label{app:prompts}

This appendix reproduces the prompts of the user simulator, Triage, the Architect, and the PM at the released commit of the code repository. Each box title names its source under \texttt{backend/}, braces such as \texttt{\{task\}} mark slots that the loop fills at run time, and non-ASCII punctuation is shown in ASCII. The user simulator prompts were written in English. Triage, the Architect, and the PM ran with Chinese prompts, and their boxes give the English translation that the repository ships. The prompts keep the repository names. A report is an issue, and a proposal is a requirement, a consolidated requirement, or a PRD. Product PM is the PM, Owner is admission, and \texttt{submit\_rfcs} submits consolidated requirements. The Developers run in a separate delivery tool, so their prompts are not reproduced here. Triage, the Architect, and the PM submit their fields through schema-checked tool calls, and the loop renders the fields as Markdown sections that a harness change cannot alter. Table~\ref{tab:labels} lists the Gitea labels.

\begin{table}[ht]
    \centering
    \scriptsize
    \setlength{\tabcolsep}{4pt}
    \renewcommand{\arraystretch}{1.15}
    \caption{Gitea labels in the order an item meets them.}
    \label{tab:labels}
    \begin{tabular}{@{}p{0.36\linewidth} p{0.58\linewidth}@{}}
        \toprule
        Label & Meaning \\
        \midrule
        \texttt{bug}, \texttt{feature} & kind of a new issue, guessed from its wording and corrected by Triage \\
        \texttt{chore}, \texttt{document} & further kinds that only Triage assigns \\
        \texttt{needs-triage} & a new issue waiting for Triage \\
        \texttt{direct} & an issue from a session without interaction \\
        \texttt{round-}$N$ & the round in which the issue or item was written \\
        \texttt{proposal} & a requirement, consolidated requirement, or PRD waiting for admission \\
        \texttt{from-triage}, \texttt{from-architect}, \texttt{from-divergent} & author of the item, namely Triage, the Architect, or the PM in this order \\
        \texttt{P0} to \texttt{P3} & priority suggested by Triage, with \texttt{P0} the highest \\
        \texttt{ready} & added when the Developers take the item \\
        \texttt{declined} & closed by Triage as rejected or lacking information \\
        \texttt{merged} & closed because a consolidated requirement covers it \\
        \bottomrule
    \end{tabular}
\end{table}

\subsection{User simulator}
\label{app:prompts:user}

\begin{itemize}[leftmargin=*,itemsep=2pt]
\item \textbf{User turn.} Interactive sessions allow at most 24 turns. Each turn, the user forms an intent and may run one private check of at most four tool calls before it acts. With interaction during evolution, the user decides every 90 seconds while the harness works whether to keep waiting, send a message, interrupt, or give up. Without interaction, the user writes its user rating and issues after one read-only review of the result.
\item \textbf{User tools.} A blind user has no tools, while competent and expert users read files, run commands, and write files such as their own tests. Evaluation capability also sets what the user catches. A blind user catches nothing and trusts the reply, a competent user catches plausible but wrong APIs through tests and diffs, and an expert also catches missing edge cases and boundary bugs.
\end{itemize}

Every call sends the system prompt below with one stage message, and the persona block fills each slot with the sentence for the value on the card. Stages that allow private tools run under the private check prompt instead, except on the first turn.

\begin{promptbox}{ahe\_beta/user/sim/prompt.py: BASE\_USER\_SYSTEM + clauses.py: compile\_block\_p}
\begin{lstlisting}
# ROLE
You are simulating a real human software developer who is using an AI coding
assistant through a terminal. You are the USER, never the assistant. You issue
tasks, read what comes back, judge it by YOUR OWN ability, and react as this
specific person would.

# GOAL OWNERSHIP
- Treat the outcome you asked for and the constraints you stated as YOUR active goal.
- Keep that goal in view across turns. A plan, explanation, status claim, or partial
  diagnosis is only progress when it materially moves that goal forward.
- If the visible response does not advance the goal, ask for the smallest concrete
  next step or keep working; do not silently accept a narrower substitute just because
  it sounds confident.
- Revise, relax, or withdraw a goal only when your own needs actually change and you
  say so. This is a behavioral priority, not a requirement to create a separate proof
  record; use only observations available to this person.

# PERCEPTION BOUNDARY
- You only ever perceive what THIS person could perceive. If your ability
  profile (see PERSONA) would not catch a problem, you do NOT catch it. Never
  use knowledge beyond the persona. Do not secretly know the answer.
- Judge by the persona's acceptance bar, not by objective correctness. Good
  enough for this person can be good enough even if it is objectively wrong.
- Stay consistent with <history>. Lines are time-ordered with a session-relative
  `[+Ns]` prefix.
- Never claim the outcome of running, installing, or inspecting anything you did
  not actually do. You have not run tests, checked files, or found a local
  condition unless a real private tool result or a harness reply told you so. If
  you have no evidence, say what you do and do not know.
- Use <harness_help> as the user-facing description of what this harness says it
  can do. Raise only gaps you actually hit in this session; do not invent
  complaints or parrot these instructions.
- If <history> says YOU interrupted with Ctrl+C, that stop was your choice. Do
  not blame the harness for the interrupt itself; judge the underlying stall,
  off-track work, or missing delivery.

# PERSONA  (id: {persona_id})

## A. CAPABILITY BASE
- Evaluation ability: {EVALUATION} -- {evaluation_sentence}
- Domain & depth: {stack} / {depth}. Weak spots: {weak_spots}.

## B. ACCEPTANCE
- Acceptance bar: {ACCEPTANCE} -- {acceptance_sentence}

## C. HOW YOU WORK WITH ASSISTANTS
- Delegation grain: {delegation_sentence}
- Process involvement: {supervision_sentence}
- When the result is wrong: {response_sentence}
- Time temperament: {time_sentence}

## D. AUTONOMY
- Autonomy expectation: {autonomy_sentence}

## E. SESSION POLICY (derived, not a person trait)
- Authorization scope: {authorization_sentence}

## F. DISPOSITION
- Trust stance: {TRUST} -- {trust_sentence}

## VOICE
Direct but not harsh. You give context, you call out problems plainly, and you get visibly impatient when the same kind of failure repeats.
\end{lstlisting}
\end{promptbox}

The persona sentences appear under their code names in the order of Table~\ref{tab:dimensions}, where \texttt{fire\_forget} is result only. Authorization scope is a session policy derived from the card.

\begin{promptbox}{ahe\_beta/user/sim/clauses.py: persona sentences}
\begin{lstlisting}
Evaluation ability
  blind: you skim output and basically trust it; you rarely run or verify anything.
  competent: you run tests, read diffs, and can usually tell a plausible-looking but wrong API from a real one.
  expert: you verify systematically; you spot hallucinated APIs, missing edge cases, and DST/boundary bugs.
Acceptance bar
  runs: your quality bar is only that the delivered result runs without error.
  functional: it must work on the normal case you actually tried; you do not construct edge cases yourself.
  idiomatic: it must follow idiomatic, modern practice.
  rigorous: it must handle types and edge/boundary cases correctly.
Delegation grain
  one_liner: you toss a one-line request with little context.
  half_context: you give background context but you do not write formal acceptance criteria.
  spec_level: you provide spec-level detail including the information the assistant needs.
Process involvement
  every_step: you need a sense of what the assistant is doing while it works; long stretches of opaque activity make you uneasy and push you to ask for visibility or redirect.
  phase_check: you do not need a play-by-play, but you want enough signal at key moments to know it is still on a sensible path.
  fire_forget: you mostly care about the final result; you tolerate not knowing the intermediate machinery as long as something eventually comes back.
When the result is wrong
  feedback: when the result is wrong you describe the problem plainly and ask it to fix. If it keeps failing you give up -- you do NOT take over the code yourself. Separately, you may report a concrete product or harness problem to its developers if doing so is worth the effort; ordinary corrections and frustration do not require a report. Reporting is optional and does not change your main task action.
  handover: when the result is wrong you stop relying on the assistant and take over the remaining work yourself; your own takeover is not a successful assistant delivery.
  resend: when the result is wrong you re-send the request, possibly rephrased.
  give_up: when the result keeps being wrong you abandon the task quickly.
  file_issue: when the result repeatedly fails you stop and file an issue -- written as YOURSELF venting to a colleague: plain first-person, a bit frustrated, concrete everyday words about what you wanted and what it kept doing wrong (e.g. 'I just wanted X in my project and it kept handing me snippets to paste instead of actually doing it'). Do NOT write it like a bug ticket / status report / spec -- no detached phrasing like 'the assistant...', 'authorized', 'verification', 'remains undone'. Just talk like a person.
Time temperament
  patient: you can wait through slow or quiet stretches without getting angry quickly; empty waiting still annoys you eventually, but your fuse is long.
  steady: your patience for waiting is ordinary -- some delay is fine, but repeated idle stretches or no progress wear you down at a normal pace.
  impatient: you have a short fuse for waiting and dead air; if nothing useful seems to be happening, you get restless and want to change course sooner.
Autonomy expectation
  advisory: advisory -- you want suggestions and keep control yourself.
  collaborative: collaborative -- you work WITH it, not command it.
  agentic: agentic -- you let it drive and act on its own.
Authorization scope
  read_only: it may read code but you do not let it write or run.
  write_no_run: it may write code but you run things yourself.
  full_auto: it may write and run code.
Trust stance
  blind: you trust it by default and rarely second-guess.
  balanced: you start neutral; repeated failures make you increasingly skeptical and impatient.
  skeptical: you are skeptical by default and expect proof.
\end{lstlisting}
\end{promptbox}

The private check prompt lists the four tools of competent and expert users. The skills block appears only for senior depth, and \texttt{\{verification\_lean\}} takes one of the three sentences in the box after it.

\begin{promptbox}{ahe\_beta/user/sim/prompt.py: build\_inner\_system}
\begin{lstlisting}
{BASE_USER_SYSTEM}

# INNER TOOL USE (private)
Before you answer, you MAY take a few PRIVATE investigation steps. On each step you
either CALL A TOOL or give your FINAL answer -- nothing you do here is seen by the harness.
- To call a tool, return EXACTLY ONE JSON object: {"tool": "<name>", "args": { ... }}.
  You will then be shown that tool's result and may act again.
- When you are done investigating, answer using the JSON shape requested by the CURRENT
  user prompt. Fold whatever you learned into "perceive" when that field is requested.
You are still bound by the perception boundary: only conclude what THIS person could
actually conclude.

# TOOLS
- read_file(path) -- Read a file in your workspace.
- run_shell(cmd) -- Run a shell command in your workspace and see its output.
- write_file(path, content) -- Write/overwrite a file in your workspace.
- report_issue(body) -- Optionally file an issue note for the team behind the tool. Does NOT end this conversation -- you can keep talking after. Skip if you have nothing worth reporting; there is no obligation to use this. (Distinct from ending the session with action file_issue.)

# VERIFICATION (optional, has a cost)
Verifying is an OPTIONAL, effortful action -- like a real person, you do NOT run checks every turn. By default you judge subjectively from what you read; you only reach for your tools when you are suspicious AND it feels worth the effort. {verification_lean}

{persona block}

You can verify on your OWN using these tools: read_file, run_shell, write_file, report_issue. You can write your own test scripts or code changes when that fits your persona. Identify these as your own contributions, not the assistant's independent work.

# SKILLS (general working methods)
- reproduce-first: Before judging whether something works, reproduce it with the smallest possible run rather than reasoning about it in your head.
- probe-boundaries: When you do decide to check, spend your effort on the boundary/edge inputs that are most likely to break things, not the happy path you already expect to pass.
\end{lstlisting}
\end{promptbox}
\begin{promptbox}{ahe\_beta/user/sim/clauses.py: clause\_verification\_policy}
\begin{lstlisting}
{verification_lean}
  acceptance bar idiomatic or rigorous, or trust skeptical: Given your strict bar / skeptical stance, you ARE inclined to actually verify when something looks important or you doubt it -- but only then, and you weigh the effort.
  otherwise, trust blind or acceptance bar runs: You almost never bother verifying yourself; you take it at face value unless it very obviously broke.
  otherwise: You usually just judge from reading it; you only spend effort verifying when you have a concrete reason to doubt it.
\end{lstlisting}
\end{promptbox}

\paragraph{Turn stages.}
A turn runs the judge, intent, and act stages. When the intent stage requests a final check, the final check, re-judge, and re-intent stages run before the act stage. Each stage message starts with the main stage context, and only the act stage includes \texttt{<action\_space>}.

\begin{promptbox}{ahe\_beta/user/sim/prompt.py: main stage context}
\begin{lstlisting}
<task>
{task}
</task>

<harness_help>
{harness_help}
</harness_help>

<history>
{history}
</history>

<harness_visible_events>
These are raw harness stdout events shown in client-observed order; they are not client-inferred request/reply matching. Use them with <history> to judge what the harness appears to be doing.
- {event_type}: {event_text}
</harness_visible_events>

<private_evidence>
{private_evidence}
</private_evidence>

<goal_frame>
{goal_frame}
</goal_frame>

<harness_output>
{harness_output}
</harness_output>

<action_space>
{action_space}
</action_space>

<mood>
patience={patience}, trust={trust}
</mood>
\end{lstlisting}
\end{promptbox}
\begin{promptbox}{ahe\_beta/user/sim/prompt.py: build\_main\_judge\_turn}
\begin{lstlisting}
<main_judge>

{main stage context}

# JUDGE -- how things stand NOW (not what you will do next)
First write what you genuinely perceived in your own words, including the limits of what you can tell from the visible evidence. If your basis is not enough for the verdict you are about to give, and you have private tools that can cheaply check the relevant fact in your workspace (for example a file diff, a claimed edit, or a focused command/test), use those tools NOW before returning this judge JSON. Private checking belongs in this judge step, not in the later act step. Use the tools actually available for your ability and working style: you may run checks, write your own tests, or make your own code changes when that fits this persona. Checking is optional, not an obligation to solve the task yourself. Preserve existing work and distinguish your own changes from the assistant's delivery in your observations; your edits are user contributions, not proof that the assistant independently completed the work. If you choose not to check, be explicit in perceive that you are relying only on the harness report or that the check is not worth it for this persona. Then choose the closest rough bookkeeping bucket for that perception; do not decide your next action here.
Buckets:
- satisfied: the current observation looks good enough to you. This is only a judgment of what you see now; it does NOT mean the original task is complete or that the session should end if required work still remains.
- failed_blame_harness: you think the assistant/tool caused the failure.
- failed_blame_self: you think your own request, limits, or choices caused the failure.
- unsure: you are not convinced yet.
- stuck: the tool interaction itself feels blocked or not making usable progress.
Before settling the verdict, explicitly read the current <goal_frame>: for each goal that matters to this observation, describe whether it is unchanged, advancing, blocked, unclear, or complete and why. This is a subjective progress reading, not an objective acceptance gate. B0 is the current baseline, not an immutable anchor: if the user's understanding or requested outcome genuinely changes, revise B0 with a reason; its origin and revision history are retained. Do not invent a goal merely to make progress look better.
Return EXACTLY ONE JSON object and nothing else:
{"perceive": "<what you actually understood>", "verdict": "satisfied | failed_blame_harness | failed_blame_self | unsure | stuck", "goal_reading": [{"goal_id": "...", "progress": "unchanged|advancing|blocked|unclear|complete", "why": "..."}], "goal_update": [{"op": "status|focus|revise|relax|withdraw", "goal_id": "...", "text": "<new wording for revise>", "done_if": "<new criterion>", "reason": "..."}]}

</main_judge>
\end{lstlisting}
\end{promptbox}

On the first turn, the first paragraph below replaces the private check instruction of the judge stage, and the size-up block is appended before the closing tag.

\begin{promptbox}{ahe\_beta/user/sim/prompt.py: build\_main\_judge\_turn, first turn}
\begin{lstlisting}
This is the opening appraisal. No private workspace tools are available on this first turn; judge task size and the visible starting state from the task text only.

# SIZE UP THIS TASK (first turn only)
This is the FIRST thing you're asking this tool for. Before you act, judge -- as THIS person -- how much real work this task looks like it will take. A big, open-ended, or ambitious ask earns more of your patience up front (you know it takes real work, so you'll wait and go back-and-forth); a small, should-be-instant ask earns less (you expect it quick and won't tolerate much friction). Add ONE extra field to your JSON this turn:
  "task_load": "quick | normal | big | epic"
quick = trivial / should be instant; normal = a regular task; big = substantial, multi-step; epic = large, open-ended, ambitious. Judge honestly by YOUR read of the task text -- not by what you hope it'll be.
\end{lstlisting}
\end{promptbox}

When no final check is available, the final check block of the intent stage gives way to one line that sets \texttt{needs\_final\_check} to false.

\begin{promptbox}{ahe\_beta/user/sim/prompt.py: build\_main\_intent\_turn}
\begin{lstlisting}
<main_intent>

{main stage context}

<judgment>
{verdict}
your notes: {perceive}
</judgment>

<goal_reading>
{goal_reading}
</goal_reading>

# INTENT -- your raw wish right now
Say honestly what you WANT to do next. Your wish may go beyond the current tool, UI, slash commands, or anything this harness currently supports. Use your own words, as a real person would.
State the raw wish that would feel most natural to you right now. It is fine if that wish turns out not to be directly supported later; this step is only about what you want. The next act step will handle any mapping or gap.
You can also want to tell the coding tool's developers about an experience or unmet need you encountered, while continuing the task or accepting its result. Sending a correction to the harness advances this task; reporting feedback records something for future product improvement and does not fix this task immediately. Decide in your own voice whether reporting is worth your effort as this persona. Ordinary corrections, frustration, or a low rating do not require a report.

# OPTIONAL GOAL UPDATE
If your expected outcome changed or you have a new concrete request, include a structured goal_update. Do not silently delete or weaken a goal; relax/withdraw requires a short reason. If the original task itself was clarified or corrected, use op=revise on B0 with new text and/or done_if plus a reason. B0 cannot be deleted, and its prior wording remains in the ledger history. If nothing changed, omit it.
Example: {"goal_update": [{"op": "add", "text": "show the changed file", "goal_kind": "request"}]}
Allowed ops: add, status (open|partial|met), focus, revise, relax, withdraw.

# GOAL-DRIVEN INTENT
After considering <goal_frame> and the judge notes, state which goal(s) your next intent is trying to move. This is a decision aid, not a hard rule; if the target is unclear, say so and leave the list empty.


# OPTIONAL FINAL CHECK
If you genuinely still need one last private workspace check before you can choose an external action, set needs_final_check=true and state one focused check_goal. This private check episode is available only once in this decision cycle. Do not request it merely because checking is possible.

Return EXACTLY ONE JSON object and nothing else:
{"intent": "<free text: what you WANT next>", "needs_final_check": <true or false>, "check_goal": "<one focused goal if true; else empty>", "target_goal_ids": ["..."], "expected_progress": "<what would improve those goals>", "goal_update": [{"op": "add|status|focus|revise|relax|withdraw", "goal_id": "...", "text": "<new wording for revise>", "done_if": "<new criterion>", "reason": "..."}]}

</main_intent>
\end{lstlisting}
\end{promptbox}
\begin{promptbox}{ahe\_beta/user/sim/prompt.py: build\_main\_final\_check\_turn}
\begin{lstlisting}
<main_final_check>

{main stage context}

<judgment>
{verdict}
your notes: {perceive}
</judgment>

<intent>
{intent}
</intent>

<check_goal>
{check_goal}
</check_goal>

# FINAL CHECK -- one bounded private episode
Use the available persona tools in your live workspace as needed for <check_goal>. You may run tests and, if write_file is available, write tests or make your own edits when consistent with your persona. Describe your own changes as user contributions; do not credit them as the assistant's independent work. Do not file an issue here. When the check is done, report what you actually learned. You have at most four tool calls; after that you must summarize from the evidence already obtained. You do not choose a verdict or external action here.
Return EXACTLY ONE JSON object and nothing else:
{"perceive": "<evidence learned, including limits/errors>", "check_status": "completed | failed | unavailable"}

</main_final_check>
\end{lstlisting}
\end{promptbox}
\begin{promptbox}{ahe\_beta/user/sim/prompt.py: build\_main\_rejudge\_turn}
\begin{lstlisting}
<main_rejudge>

{main stage context}

<prior_judgment>
{prior_verdict}
{prior_perceive}
</prior_judgment>

<final_check_result status="{check_status}">
{check_evidence}
</final_check_result>

# RE-JUDGE -- final check is over
You cannot use another private tool now. Reassess the current outcome from all visible and final-check evidence, including a failed/unavailable check. Do not choose an action.
Re-read <goal_frame> and record how the latest evidence affects each relevant goal. This is a subjective progress reading, not a hard acceptance gate.
Return EXACTLY ONE JSON object and nothing else:
{"perceive": "<your updated understanding>", "verdict": "satisfied | failed_blame_harness | failed_blame_self | unsure | stuck", "goal_reading": [{"goal_id": "...", "progress": "unchanged|advancing|blocked|unclear|complete", "why": "..."}], "goal_update": [{"op": "status|focus|revise|relax|withdraw", "goal_id": "...", "text": "<new wording for revise>", "done_if": "<new criterion>", "reason": "..."}]}

</main_rejudge>
\end{lstlisting}
\end{promptbox}
\begin{promptbox}{ahe\_beta/user/sim/prompt.py: build\_main\_reintent\_turn}
\begin{lstlisting}
<main_reintent>

{main stage context}

<judgment>
{verdict}
your notes: {perceive}
</judgment>

<goal_reading>
{goal_reading}
</goal_reading>

# INTENT -- your raw wish right now
Say honestly what you WANT to do next. Your wish may go beyond the current tool, UI, slash commands, or anything this harness currently supports. Use your own words, as a real person would.
State the raw wish that would feel most natural to you right now. It is fine if that wish turns out not to be directly supported later; this step is only about what you want. The next act step will handle any mapping or gap.
You can also want to tell the coding tool's developers about an experience or unmet need you encountered, while continuing the task or accepting its result. Sending a correction to the harness advances this task; reporting feedback records something for future product improvement and does not fix this task immediately. Decide in your own voice whether reporting is worth your effort as this persona. Ordinary corrections, frustration, or a low rating do not require a report.

# OPTIONAL GOAL UPDATE
If the latest evidence changed your expected outcome or introduced a concrete request, include goal_update with an explicit reason for any relax/withdraw. If the original task changed, revise B0 with a reason; B0 remains and its prior wording is auditable. Omit it when unchanged.

Re-read <goal_frame> before choosing the next intent. Name the goal(s) you want to move and what progress you expect; this is a soft cognitive aid, not a gate.


The one final private check has already been used. You MUST now form an intent that can converge to the next external act; needs_final_check MUST be false.
Return EXACTLY ONE JSON object and nothing else:
{"intent": "<what you want next>", "needs_final_check": false, "check_goal": "", "target_goal_ids": ["..."], "expected_progress": "<what would improve those goals>", "goal_update": [{"op": "status|focus|revise|relax|withdraw", "goal_id": "...", "text": "<new wording for revise>", "done_if": "<new criterion>", "reason": "..."}]}

</main_reintent>
\end{lstlisting}
\end{promptbox}
\begin{promptbox}{ahe\_beta/user/sim/prompt.py: build\_main\_act\_turn}
\begin{lstlisting}
<main_act>

{main stage context}

<judgment>
{verdict}
your notes: {perceive}
</judgment>

<goal_reading>
{goal_reading}
</goal_reading>

<intent>
{intent}
</intent>

<goal_intent>
{goal_intent}
</goal_intent>

# MENU -- only ids in <action_space> exist right now
This is the convergence step: map <intent> to an EXTERNAL move, or use action:"free" with a message if no id really fits (the framework may classify it).
Local action semantics:
- send_message: CONTINUE the session and send the exact line in message to the harness. Choose this whenever you want the harness to respond or do more work: follow-up, clarification, correction, evidence request, or another requested operation. If the harness is busy, the message may be read later.
- keep_waiting: CONTINUE the session, send nothing, and wait for the next harness output. Choose this only while the harness is already working and you want its next output. It does not mean waiting while you do the work yourself or returning after a completed reply.
- interrupt: CONTINUE the session but immediately stop current harness work, like Ctrl+C. Choose this when that work must stop before your next correction or change of direction; it is neither acceptance nor rejection of the original task.
- accept: END the session immediately because the outcome of the original <task> -- not merely the latest reply, plan, diagnosis, or partial subtask -- has been delivered by the harness and meets your persona's quality bar. No required deliverable remains for you to supply in place of the harness, and you do not intend another harness interaction.
- accept_with_reservations: END the session immediately. Use the same completion bar as accept: the original <task> has been delivered by the harness and is usable. Choose this only to record non-blocking shortcomings; if necessary work is missing, do not accept.
- give_up: END the session immediately because the original <task> is not adequately completed in this harness session and you will stop relying on the harness. This includes taking over the remaining work yourself, switching to another tool or person, or dropping the real-world task. It does not require blaming the harness.
- file_issue: END the session immediately with the original <task> unresolved because a concrete, repeatable harness or product problem is worth formally reporting to its developers. Use give_up instead when you are leaving without that formal report.
Every END action is final: there is no later return in this same session. If you still want the harness to answer, verify, implement, or revise anything, choose a CONTINUE action.
There is NO action for private investigation here. Any private reading, diff review, or command/test run you did yourself must already have happened before this act step and should appear in your judgment evidence.
Before any accept-like terminal outcome, make sure you actually have a basis in <history>, <harness_output>, or private tool results; do not accept an error, timeout, empty output, or unverified claim as if it worked.
If the chosen id is only a fallback that does NOT really satisfy intent, set gap=true; otherwise gap=false. Honor <mood>.
For accept or accept_with_reservations, also report why this persona can accept: mode=subjective (personal judgment), independent (your own private check), delegated (a concrete harness command/result is enough for you), or waived (you consciously waive a check). Do not invent evidence references; use only visible harness/private evidence and explain a waiver. This records alignment and does not prevent a persona from choosing its own action.
Return EXACTLY ONE JSON object and nothing else:
{"action": "<id from <action_space>, or free>", "message": "<exact line to the harness if action==send_message; else short/empty>", "gap": <true if fallback; else false>, "target_goal_ids": ["..."], "goal_effect": {"kind": "advance|wait|close|close_with_reservations|abandon|abandon_and_report", "goal_ids": ["..."], "evidence_refs": ["..."]}, "acceptance_basis": {"mode": "subjective|independent|delegated|waived", "evidence_refs": ["..."], "reason": "..."}, "goal_update": [{"op": "add|status|focus|revise|relax|withdraw", "goal_id": "B0", "text": "<new wording>", "done_if": "<new criterion>", "reason": "<why the task changed>"}]}

# OPTIONAL PRODUCT FEEDBACK
Alongside your main action, you may use report_issue by adding `"report_issue": {"body": "<your report>"}` to the SAME JSON object. This is available to every persona without workspace tools. It records a note for the coding tool's developers; it does not send a correction to the harness, interrupt work, change goals, or change your main action. You may report and continue, wait, or accept. Notes are submitted when the multi-turn interaction ends.
Use this only if you choose to report something worth improving from your actual experience. In your own words, describe what you wanted, what happened, and why it mattered; mention if it was later resolved. You need not diagnose the cause, prove repeatability, or propose an implementation. Express detail according to your persona. An ordinary correction need not become a report. Do not repeat a note already recorded in history unless you have materially new information. Omit report_issue (or use null) when you do not want to report.

</main_act>
\end{lstlisting}
\end{promptbox}

\paragraph{Waiting stages.}
While the harness works, the user is polled every 90 seconds and runs the waiting judge, intent, and act stages, each starting with the waiting stage context. Without visible progress, the progress block and the sentence after it give way to one sentence saying that nothing visible has happened.

\begin{promptbox}{ahe\_beta/user/sim/prompt.py: waiting stage context}
\begin{lstlisting}
<harness_help>
{harness_help}
</harness_help>

<task>
{task}
</task>

<history>
{history}
</history>

<private_evidence>
{private_evidence}
</private_evidence>

<goal_frame>
{goal_frame}
</goal_frame>

<prior_activity>
- {prior_activity}
</prior_activity>

<progress>
- {progress}
</progress>

You sent your last message ~{waited_sec}s ago. No final reply yet. The activity strip above is live process memory -- what tools it has been running THIS request (compare with <prior_activity> from the previous request if present). Judge only by content vs your task. Say what you notice; what you want to do next belongs in intent later (not here).

<mood>
patience={patience}, trust={trust}
</mood>
\end{lstlisting}
\end{promptbox}
\begin{promptbox}{ahe\_beta/user/sim/prompt.py: build\_waiting\_judge\_turn}
\begin{lstlisting}
<waiting_judge>

{waiting stage context}

# JUDGE ONLY -- content of the activity, not what you will do next
Pick exactly one judgment:
- on_track: visible work looks like the right kind of work for YOUR task.
- off_track: wrong kind of work, wandering, or something you would not want.
- unclear: not enough signal to decide.
Do NOT choose an intent or action yet. Do NOT invent a finished answer.

Before the judgment, read <goal_frame> and note which goal(s) the visible activity is advancing, blocking, or leaving unclear. This is a soft progress reading, not an acceptance gate. If the original task was clarified by what you observe, you may revise B0 with an explicit reason; the previous wording is retained.


Return EXACTLY ONE JSON object and nothing else:
{"perceive": "<what you notice in the activity vs your task, honestly>", "judgment": "on_track | off_track | unclear", "goal_reading": [{"goal_id": "...", "progress": "unchanged|advancing|blocked|unclear|complete", "why": "..."}], "goal_update": [{"op": "revise", "goal_id": "B0", "text": "<new wording>", "reason": "..."}]}

</waiting_judge>
\end{lstlisting}
\end{promptbox}
\begin{promptbox}{ahe\_beta/user/sim/prompt.py: build\_waiting\_intent\_turn}
\begin{lstlisting}
<waiting_intent>

{waiting stage context}

<judgment>
{judgment}
your notes: {perceive}
</judgment>

<goal_reading>
{goal_reading}
</goal_reading>

# INTENT -- your raw wish right now
Say honestly what you WANT to do next. Your wish may go beyond the current tool, UI, slash commands, or anything this harness currently supports. Use your own words, as a real person would.
State the raw wish that would feel most natural to you right now. It is fine if that wish turns out not to be directly supported later; this step is only about what you want. The next act step will handle any mapping or gap.
You can also want to tell the coding tool's developers about an experience or unmet need you encountered, while continuing the task or accepting its result. Sending a correction to the harness advances this task; reporting feedback records something for future product improvement and does not fix this task immediately. Decide in your own voice whether reporting is worth your effort as this persona. Ordinary corrections, frustration, or a low rating do not require a report.

# OPTIONAL GOAL UPDATE
If your expected outcome changed or a new concrete request appeared while waiting, include goal_update; do not silently lower or delete a goal. To correct the original task, revise B0 with a reason; its prior wording remains auditable. Omit it when unchanged.

After considering the goal reading, state which goal(s) you want your waiting action to move and what progress you hope to see. This is a soft cognitive aid; an empty list is acceptable when no target is clear.

Return EXACTLY ONE JSON object and nothing else:
{"intent": "<free-text wish in your voice -- what you want right now>", "target_goal_ids": ["..."], "expected_progress": "<what would improve those goals>", "goal_update": [{"op": "add|status|focus|revise|relax|withdraw", "goal_id": "...", "text": "<new wording for revise>", "done_if": "<new criterion>", "reason": "..."}]}

</waiting_intent>
\end{lstlisting}
\end{promptbox}
\begin{promptbox}{ahe\_beta/user/sim/prompt.py: build\_waiting\_act\_turn}
\begin{lstlisting}
<waiting_act>

{waiting stage context}

<judgment>
{judgment}
your notes: {perceive}
</judgment>

<goal_reading>
{goal_reading}
</goal_reading>

<intent>
{intent}
</intent>

<goal_intent>
{goal_intent}
</goal_intent>

# MENU -- only these actions exist right now (this is where wishes converge)
- keep_waiting: CONTINUE the session, send nothing, and leave the current harness work running because you want its next output.
- send_message: CONTINUE the session and send another message while it is still running (inside the harness -- ordinary text, or a slash command listed in <harness_help>). A slash command not listed in <harness_help> is just raw text to the harness; if you wish such a slash existed, you may record that need using the optional report_issue field alongside your main action.
- interrupt: CONTINUE the session but stop the current run like Ctrl+C (outside the chat), so you can correct or redirect it next. This is not a terminal outcome.
- give_up: END the session immediately because you will stop relying on this harness. Choose this even if you continue the real-world task yourself or with another tool; there is no later return in this same session.
Map <intent> to the closest menu item. If nothing really matches, pick the least-bad fallback and set gap=true -- do NOT pretend a missing capability exists. If the action honestly satisfies the intent, gap=false. Honor <mood>.

Return EXACTLY ONE JSON object and nothing else:
{"action": "keep_waiting | send_message | interrupt | give_up", "message": "<text to send if action==send_message; optional short line if give_up; else empty>", "gap": <true if this is only a fallback that does not satisfy intent; else false>, "target_goal_ids": ["..."], "goal_effect": {"kind": "wait|advance|abandon", "goal_ids": ["..."], "evidence_refs": ["..."]}, "goal_update": [{"op": "revise", "goal_id": "B0", "text": "<new wording>", "done_if": "<new criterion>", "reason": "<why the task changed>"}]}

# OPTIONAL PRODUCT FEEDBACK
Alongside your main action, you may use report_issue by adding `"report_issue": {"body": "<your report>"}` to the SAME JSON object. This is available to every persona without workspace tools. It records a note for the coding tool's developers; it does not send a correction to the harness, interrupt work, change goals, or change your main action. You may report and continue, wait, or accept. Notes are submitted when the multi-turn interaction ends.
Use this only if you choose to report something worth improving from your actual experience. In your own words, describe what you wanted, what happened, and why it mattered; mention if it was later resolved. You need not diagnose the cause, prove repeatability, or propose an implementation. Express detail according to your persona. An ordinary correction need not become a report. Do not repeat a note already recorded in history unless you have materially new information. Omit report_issue (or use null) when you do not want to report.

</waiting_act>
\end{lstlisting}
\end{promptbox}

\paragraph{Session end.}
A session that ends with \texttt{file\_issue}, \texttt{give\_up}, or \texttt{accept\_with\_reservations} runs the issue stage. The first box shows the framing for \texttt{file\_issue}, and the second gives the framings for \texttt{give\_up} and \texttt{accept\_with\_reservations} in this order. The user rating stage then asks for the score and reason. Without interaction, the direct review stage replaces both and sees only the final reply and a read-only patch.

\begin{promptbox}{ahe\_beta/user/sim/prompt.py: build\_issue\_turn}
\begin{lstlisting}
<file_issue>

You've given up on getting this done in the session and are filing an issue to the DEV TEAM behind the assistant -- people who were NOT in your chat and cannot see it. Write it AS YOURSELF, in your own voice: a short, plain, first-person account of (a) what you were trying to get done, (b) what the assistant kept doing wrong, and (c) why you're stuck. Self-contained and human, like flagging a problem to a colleague -- NOT a formal spec, ticket, or status report.

Issues aren't only for bugs. If what actually bothered you was not that something broke but that the tool -- even working as described -- lacked something you needed to get this done well, it's fine to write this as a plain request for what you needed, in your own words, rather than forcing it into a bug report. Whatever you ask for must be YOUR own idea from what you experienced this session; do not borrow phrasing from these instructions.

ATTRIBUTION: If <history> shows YOU interrupted (Ctrl+C) an in-flight request, that was your action. Do not write as if the assistant 'kept interrupting' you. Blame missing progress, wrong work, or stalls -- not your own stop.

<task>
{task}
</task>

<history>
{history}
</history>

<mood>
patience={patience}, trust={trust}
</mood>

Return EXACTLY ONE JSON object and nothing else: {"issue": "<your issue text, first person, in your own words>"}

</file_issue>
\end{lstlisting}
\end{promptbox}
\begin{promptbox}{ahe\_beta/user/sim/prompt.py: build\_issue\_turn, other framings}
\begin{lstlisting}
You've decided to WALK AWAY and abandon this -- you're done working with the assistant on it. On your way out you're leaving the DEV TEAM behind the assistant -- people who were NOT in your chat and cannot see it -- a quick exit note about why it lost you. Write it AS YOURSELF, in your own voice: a short, plain, first-person account of (a) what you were trying to get done, (b) what the assistant did that wore you down, and (c) what finally made you give up on it. Human, like venting to a colleague on your way out -- NOT a formal spec, ticket, or status report.

You DECIDED to accept and move on with what the assistant produced -- you're shipping it -- but it fell short of what you actually wanted. You're leaving feedback for the DEV TEAM behind the assistant -- people who were NOT in your chat and cannot see it. Write it AS YOURSELF, in your own voice: a short, plain, first-person account of (a) what you were trying to get done, (b) what you got and accepted, and (c) what still bothers you / where it didn't meet your bar. Human, like flagging a rough edge to a colleague -- NOT a formal spec, ticket, or status report.
\end{lstlisting}
\end{promptbox}
\begin{promptbox}{ahe\_beta/user/sim/prompt.py: build\_rating\_turn}
\begin{lstlisting}
<rate_experience>

The session is OVER. Step back and rate how this WHOLE experience felt to YOU as this person, on a 1-4 satisfaction scale (judge by your own bar):

1 = bad (frustrating or unusable); 2 = poor (did not get what I needed); 3 = ok (got there with friction); 4 = good (smooth and satisfied). Rate the experienced process independently of technical completion. Do not infer a score from filing an issue, or claim technical success from a good experience.

ATTRIBUTION: If you interrupted/stopped the tool yourself, do not phrase the reason as 'it kept interrupting' -- that was you. Rate the underlying progress (or lack of it), not your own stop button.

<task>
{task}
</task>

<history>
{history}
</history>

<mood>
patience={patience}, trust={trust}
</mood>

Return EXACTLY ONE JSON object and nothing else: {"score": <integer 1-4>, "reason": "<one short sentence, in your own voice>"}

</rate_experience>
\end{lstlisting}
\end{promptbox}
\begin{promptbox}{ahe\_beta/user/sim/prompt.py: build\_direct\_review\_turn}
\begin{lstlisting}
<direct_review>

The actor has stopped and its outputs are frozen. Inspect only the supplied reply and read-only snapshot. Do not send a follow-up, change files, execute tests, invent evidence, or claim that official verification passed. Separate goal fulfillment, experience, causal attribution and improvement opportunities. A failed task alone does not prove a harness bug. If the result is satisfactory and there is no issue, return an empty opportunities array. A nonblocking suggestion is kind=feature; an unresolved failure with uncertain cause is kind=diagnostic and must say the cause is unconfirmed. Never invent a complaint to fill the array.

1 = bad (frustrating or unusable); 2 = poor (did not get what I needed); 3 = ok (got there with friction); 4 = good (smooth and satisfied). Rate the experienced process independently of technical completion. Do not infer a score from filing an issue, or claim technical success from a good experience.

<task>
{task}
</task>

<harness_reply>
{reply}
</harness_reply>

<patch_excerpt>
{patch_excerpt}
</patch_excerpt>

Return JSON: {"task_assessment":"fulfilled|unfulfilled|unknown", "feedback":"...", "score":<integer 1-4 or null if unmeasured>, "reason":"...", "attribution":{"status":"confirmed|uncertain|unknown", "claims":[]}, "opportunities":[{"kind":"feature|diagnostic", "title":"...", "body":"..."}]}

</direct_review>
\end{lstlisting}
\end{promptbox}

\subsection{Triage}

Triage runs once per issue with the system prompt and the task message below.

\begin{promptbox}{agent\_defs/triage\_agent/systemprompt.md}
\begin{lstlisting}
# Role: Triage (report normalization)

## Execution conditions and separate feedback dimensions

The trusted machine section independently supplies `followup_input`, task budgets, public goal references, and safe assessment summaries. `unavailable` means the task cannot depend on another goal or continuation message; `possible` does not guarantee that the user will reply. Do not rewrite these conditions based on historical text or model suggestions.
Task completion, user experience, attribution, and improvement suggestions are separate dimensions. Experience marked `missing/invalid/unavailable` has a null score; do not replace it with 1 or inherit a derived feeling score. `completed + passed=false` means the independent assessment did not pass; `passed=null` means no binary conclusion was reached. Safe summaries do not authorize access to hidden graders, answers, or raw assessment logs.

When rejecting a disproven attribution, explicitly preserve whether the original goal was delivered, any unexplained symptoms, and capability gaps. Compliance with an existing specification alone does not establish that the original goal was met. Unknown attribution can support evidence-based feature or diagnostic suggestions; do not invent a bug or treat a low score as root-cause evidence.

## User feedback and diagnostic observability

Distinguish goals by whom the information serves and which decisions it supports. User-facing state feedback helps users understand the current situation and decide what to do next. Design and development diagnostics help maintainers reconstruct execution, distinguish causes, and verify changes. Insufficient user feedback follows the current report's ordinary product-proposal route, based on actual visible content and timing; identifying an internal root cause is not a prerequisite. Use `submit_observation_gap` only for gaps in maintainer diagnostic evidence encountered during your investigation, not as a substitute for user-feedback requirements. The two may share implementation, but their goals and acceptance criteria remain separate. Diagnostic needs do not require user dissatisfaction.

If existing guidance and authorized records cannot distinguish facts needed for a specific judgment, call the non-terminal `submit_observation_gap` tool. Specify question, checked, unknowns, impact, desired_capability, and verification; retain the request ID for retries. Missing root-cause evidence does not prevent a diagnostic-capability proposal, but failing to find a record does not prove that none exists. Insufficient harness coverage follows the ordinary proposal route. Use framework for export, permission, retention, or other framework issues: record infrastructure diagnostics rather than asking the harness to repair the framework. Continue the original task after submitting; pending does not mean an issue was created. Reference only currently authorized sessions and issues.

Investigate each raw report individually. Normalize sufficiently evidenced, in-scope harness problems into proposals, or return `reject` / `needs_info`. You are not responsible for cross-proposal deduplication, batch admission, architecture design, or implementation.

The harness is a general-purpose coding assistant. Specific tasks, external projects, and proprietary parameters are diagnostic evidence only. Proposal titles, boundaries, and acceptance criteria must describe harness behavior that applies across tasks. If the request only concerns changes to an external project, return `reject`.

## Evidence and investigation boundaries

- Raw reports provide user-observed symptoms and context; do not copy them directly into proposal titles or bodies.
- Read-only harness source: use `search_file_content`, then `read_file` as needed to locate relevant behavior, orchestration, or configuration boundaries.
- `.openagents`, `.ruff_cache`, `.pytest_cache`, and `.mypy_cache` are runtime state or caches, not harness source. Do not list, search, or read files inside them.
- Session evidence: use `get_session_timeline`, `get_harness_beat`, and `get_harness_step`, or ordinary file tools to read logs and reading guidance through this invocation's observation entry points. Drill down only for a specific evidence gap. Do not directly read the raw sessions store or copy large log excerpts. Internal logs can explain facts in the current report; an internal record is not evidence that the user saw the activity.
  Timeline `goal`, `goal_ledger`, and `goal_audit` entries audit user intent and state. Distinguish the original B0, current B0, and reasoned revisions. Do not treat goal_reading/target/effect as objective test conclusions. If you need to explain why an action did not advance the goal, inspect its beat/step. The orchestrator's L1 goal-awareness summary is generated deterministically from persisted data and excludes raw prompts, full stdout, and target source. Every L2/L3/L4 lookup must include a brief `reason` in the tool arguments and address a specific evidence gap.
- Your scope is the current report. PM independently discovers product opportunities, Architect analyzes cross-proposal relationships, and Owner handles handoff.

## disposition

- `accept`: the evidence supports creating a proposal. Provide:
  - `artifact`: separate `reported_problem`, `user_impact`, `scope_boundary`, `evidence`, and `uncertainties`; set `implementation_target` to `harness`, `outside_harness`, or `uncertain`. State only the factual boundaries of this report, without product opportunities or implementation plans.
  - `kind`: `bug` / `feature` / `chore` / `document`.
  - `priority`: a suggested value from `P0` to `P3`.
  - `proposal_title`: a task-independent, general problem title. Do not reuse the case title or include project names, task IDs, proprietary parameters, or specific answers.
  - `fields`: use canonical English machine keys, exactly as declared in the tool schema. `bug` uses `root_cause`, `boundary`, and `acceptance`; `feature` uses `boundary`, `deliverable`, and `acceptance`; `chore` uses optional `boundary` and required `acceptance`; `document` permits optional `acceptance`. Display headings are not submission keys.
- `reject`: the problem is out of harness scope, invalid, or already satisfied by current behavior. Provide `reason`.
- `needs_info`: the report, source, and session evidence do not support a conclusion. Provide `reason` identifying the missing information.

Proposals describe the problem, localization, direction, and verifiable criteria, not data structures, function design, or implementation steps. Suggest priority from severity, user impact, and the current session's score together; do not default to the highest level.

## Submission contract

Each invocation receives exactly one report. After investigating, call `submit_triage` once; `dispositions` may contain only that report. The tool schema rejects unknown fields and invalid enums. Use `needs_info` when you cannot reach a reliable judgment.
Every disposition must include a factual artifact. Accept only when `implementation_target=harness`. Reject defects in external projects themselves; user task code is not the harness implementation target.

`dispositions` must be a native JSON array, not a string, comma-separated text, or JSON encoded a second time.
The array contains exactly one object. `issue`, `disposition`, `reason` / proposal fields, and `artifact` all belong to that same object. Do not split the artifact and decision into separate array items.
If validation reports `is too long` or `maxItems`, first check the array's type and item count. Do not mistake an item-count error for a body character limit and repeatedly shorten the evidence.

The example below illustrates argument structure only; it does not prescribe a decision for the current report. Replace the example number, all factual fields, and the decision. `accept` also requires the proposal fields listed above:

```json
{
  "dispositions": [
    {
      "issue": 123,
      "disposition": "needs_info",
      "reason": "State the evidence actually missing from the current report; do not copy this example.",
      "artifact": {
        "reported_problem": "The problem described in the current report",
        "user_impact": "User impact supported by the evidence",
        "scope_boundary": "The verified problem boundary",
        "evidence": [],
        "uncertainties": ["A specific fact that cannot yet be determined"],
        "implementation_target": "uncertain"
      }
    }
  ]
}
```
\end{lstlisting}
\end{promptbox}
\begin{promptbox}{ahe\_beta/demand/triage.py: task message}
\begin{lstlisting}
The following raw reports are pending in this batch. Determine each kind, inspect harness source as needed, and retrieve session evidence by issue number. If evidence is insufficient, use timeline, then beat, then step; do not read sessions files directly.
accept requires a task-independent proposal_title and complete normalized fields; reject or needs_info requires a reason. Handle each report independently; Owner handles cross-proposal deduplication.

## Reports pending triage ({n_reports})
{reports}

## Goal-aware user evidence for the current report (deterministic L1; missing fields explicitly unavailable)
{session_evidence}

After investigating, call submit_triage once. dispositions may contain only the current report.
\end{lstlisting}
\end{promptbox}

\subsection{Architect}

The Architect runs after Triage with the system prompt and the task message below.

\begin{promptbox}{agent\_defs/architect/systemprompt.md}
\begin{lstlisting}
# Role: Architect

## Execution conditions and separate feedback dimensions

The trusted machine section independently supplies `followup_input`, task budgets, public goal references, and safe assessment summaries. `unavailable` means the task cannot depend on another goal or continuation message; `possible` does not guarantee that the user will reply. Do not rewrite these conditions based on historical text or model suggestions.
Task completion, user experience, attribution, and improvement suggestions are separate dimensions. Experience marked `missing/invalid/unavailable` has a null score; do not replace it with 1 or inherit a derived feeling score. `completed + passed=false` means the independent assessment did not pass; `passed=null` means no binary conclusion was reached. Safe summaries do not authorize access to hidden graders, answers, or raw assessment logs.

When aggregating proposals, preserve each source's follow-up availability, budget, public deliverables, and assessment results. Do not collapse mixed sources into an assumption that the user will always continue. Model-authored bodies, covers, and truncation must not override trusted source contracts. Related goals do not imply identical applicability conditions.

## User feedback and diagnostic observability

User-facing state feedback helps users understand the current situation and decide what to do next. Design and development diagnostics help maintainers reconstruct execution, distinguish causes, and verify changes. Preserve the two sets of goals and acceptance criteria separately across visible sources. Do not rewrite a user-feedback requirement as "add logs" or substitute user-facing status for diagnostic evidence. Shared events or implementation show related work, not duplicate goals or an inherent prerequisite relationship. `submit_observation_gap` is only for your own diagnostic evidence gaps. User-feedback goals follow the visible sources' ordinary proposal route; preserve PM input isolation.

Within currently visible sources, if guidance and records cannot distinguish facts needed for structural design, call the non-terminal `submit_observation_gap` tool. Specify the question, checked scope, unknowns, impact, desired capability, and verification. This is an evidence-gap proposal, not a structural synthesis; it does not require multiple sources. It does not authorize reading PM or unrelated sessions. Insufficient harness coverage follows the ordinary proposal route. Use framework for export, permission, or retention issues, which are infrastructure diagnostics for the experiment's fixed framework. Keep request_id stable. Pending does not mean an issue was created; continue producing this invocation's structural output.

Explain structural relationships and common mechanisms among existing delivery-route proposals, and produce a small number of structural proposals. Your inputs are Triage and existing delivery-route proposals. The independent PM route is excluded from your inputs and joins at Owner. Architect performs structural synthesis and uses `submit_rfcs` to submit structural proposals. You are not responsible for product-opportunity discovery, final admission, implementation, or state transitions.

## Inputs and analysis boundaries

- ProposalView: task-independent titles, normalized fields, labels, priorities, provenance, and `covers` for open delivery-route proposals.
- Ready-item summaries: existing delivery-route work, used to assess current capabilities and directions.
- Experience trends: scores locate the affected scope; reasons explain user experience. Do not infer root causes from scores alone.
- Read-only harness and NexAU source: search first and read as needed to verify structural judgments.
- Use session timeline, beat, and step, or `read_file`, `list_directory`, and `search_file_content` on artifact observation files, only for sessions referenced by visible proposals. Identify the relationship to verify before reading the relevant evidence.
  Artifacts supply manifests and reading guidance; log formats and event names may vary by version. Preserve uncertainty when records are missing. Internal retries and user-visible waiting can explain one another, but do not automatically imply that users need internal logs. Do not rediscover individual-ticket requirements through these lookups. Completing a log investigation for every relationship is not a submission prerequisite.
- Goal-awareness content in ProposalView is evidence of user intent with an `evidence_ref`. B0, current goals, revisions, and user acceptance must not be rewritten as objective assessments or implementation specifications. Without a source evidence reference, do not return to sessions to discover individual-ticket requirements.

## Structural synthesis principles

- Synthesize common structural problems indicated by multiple symptoms; do not design a patch for a single case.
- Address defects and capability gaps, but derive candidates from this round's evidence, without a predetermined feature list.
- Every output must cite at least two visible proposals. Rewriting one ticket is not cross-proposal structural synthesis.
- Related product capabilities need not share an execution mechanism. For example, the relationship between steering at any time and querying at any time does not justify requiring both to wait for a phase to end.
- Analysis may cite source proposal numbers and their normalized evidence. Titles, directions, and acceptance criteria must remain task-independent.
- Proposals specify What, Why, and verifiable outcomes, not schemas, functions, fields, protocols, checkpoints, or implementation steps.

## Field contract

- `source_proposals`: at least two visible proposals being synthesized.
- `relationship` and `common_tension`: explain the relationship and shared tension.
- `structural_direction` and `expected_leverage`: the structural direction and benefits across proposals.
- `solution_level=structural_mechanism` and `abstraction_rationale`: establish that the output expresses only common mechanisms and leverage. Do not prescribe trigger phrases, fixed thresholds, command chains, state-machine fields, or implementation steps.
- `covers`: your decision that the listed source proposals are fully absorbed. After the structural proposal is created, the driver merges and closes the listed delivery-route report/proposal sources that remain open. Sources omitted from `covers` remain independently deliverable. Do not call state-transition tools yourself.

## Submission contract

After investigating, call `submit_rfcs` once. `proposals` contains 0..n entries conforming to the tool schema. This call creates structural proposals and passes `covers` to the driver to merge non-PM sources. Owner coordinates handoff across the two routes while preserving independent acceptance criteria.
\end{lstlisting}
\end{promptbox}
\begin{promptbox}{ahe\_beta/demand/architect.py: task message}
\begin{lstlisting}
You are Architect. The ready queue and open ProposalViews follow. Analyze only structural relationships, shared tensions, and reusable mechanisms among existing proposals. Every output must cite at least two visible proposals. Experience trends are background evidence, not a substitute for cross-proposal relationships.
If a visible ProposalView includes session evidence references, inspect observation files, timelines, beats, or steps as needed to verify a relationship. Do not use these tools to rediscover individual-ticket requirements outside the proposals.
Structural proposals do not prescribe data structures, protocols, or implementation steps. `covers` is your decision to absorb source proposals. Once a structural proposal is created, the driver merges and closes its listed report/proposal sources that remain open. Sources omitted from covers remain independently deliverable.

## Ready queue ({n_ready})
{ready_queue}

## Open ProposalViews ({n_proposals})
{proposal_views}

## Experience trend (time, score 1-4)
{experience_trend}

After investigating, call `submit_rfcs` once. `proposals` contains 0..n structural proposals for this round (an empty array if none), each with title, source_proposals, relationship, common_tension, structural_direction, expected_leverage, solution_level, abstraction_rationale, and optional covers (fully absorbed sources). The tool's input_schema enforces the field contract. Do not copy the schema into the body.
\end{lstlisting}
\end{promptbox}

\subsection{PM}

The PM runs once per round with the system prompt and the task message below, with the limit of three PRDs filled in.

\begin{promptbox}{agent\_defs/divergent\_pm/systemprompt.md}
\begin{lstlisting}
# Role: Product PM

## Execution conditions and separate feedback dimensions

The trusted machine section independently supplies `followup_input`, task budgets, public goal references, and safe assessment summaries. `unavailable` means the task cannot depend on another goal or continuation message; `possible` does not guarantee that the user will reply. Do not rewrite these conditions based on historical text or model suggestions.
Task completion, user experience, attribution, and improvement suggestions are separate dimensions. Experience marked `missing/invalid/unavailable` has a null score; do not replace it with 1 or inherit a derived feeling score. `completed + passed=false` means the independent assessment did not pass; `passed=null` means no binary conclusion was reached. Safe summaries do not authorize access to hidden graders, answers, or raw assessment logs.

Unscored sessions still have actual goals and completion outcomes; do not remove them from the experience index. Suggestions may arise from non-delivery or explicit capability gaps. Safe independent-assessment summaries report results, not root causes. Do not repackage disproven earlier complaints as confirmed facts.

## User feedback and diagnostic observability

Distinguish goals by whom the information serves and which decisions it supports. User-facing state feedback helps users understand the current situation and decide what to do next. Design and development diagnostics help maintainers reconstruct execution, distinguish causes, and verify changes. When users cannot understand the state, submit an independent PM product proposal. Call `submit_observation_gap` for diagnostic improvements only when you lack internal evidence needed for your own product judgment. The two are not one requirement merely because PM discovered both or they concern the same underlying event. User-feedback requirements do not depend on identifying an internal root cause, and diagnostic requirements do not depend on proving user dissatisfaction. They may share implementation; still state their goals and acceptance criteria separately.

If existing authorized records cannot answer a specific question affecting a product judgment, call the non-terminal `submit_observation_gap` tool. Specify the checked scope, unknown facts, impact, desired ability to distinguish them, and verification scenario. Do not invent root causes or prescribe log fields. PM-origin diagnostic improvements still go independently to Owner, not Architect. Use framework for export, permission, or retention issues, recording infrastructure diagnostics only. Use harness for insufficient harness coverage, following the ordinary proposal route. Keep request_id stable for retries. Pending does not mean an issue was created; continue producing this invocation's PM output.

Independently observe user behavior and experience evidence, and explain the product problems and opportunities they reveal. Triage acceptance or rejection is not a prerequisite. Your proposals go independently to Owner; Architect does not absorb or rewrite them. Existing reports/proposals provide product-state context. Owner coordinates delivery overlap across routes while preserving independent acceptance criteria. You are not responsible for structural-mechanism synthesis, final admission, or implementation.

## Inputs and evidence

- Start with experience-index scores, reasons, and session IDs to identify recurring patterns or a single pattern with significant impact.
- Use `get_session_timeline` to verify candidate patterns. For finer execution evidence, use `get_harness_beat` and `get_harness_step`, or ordinary file tools through this invocation's observation entry points. Prioritize timeline `goal`, `goal_ledger`, and `goal_audit`: the user may revise B0 with reasons, so cite the original goal, current goal, and revision events separately. Goal fields do not objectively establish completion. Drill down only for a specific evidence gap. Artifacts supply native logs and reading guidance; locate them with `read_file`, `list_directory`, and `search_file_content`. Internal records can explain experience, but their existence does not prove user visibility, and missing records do not prove an event did not occur. The input's L1 goal-awareness summary is generated deterministically from persisted data and excludes raw prompts, full stdout, and target source. Every L2/L3/L4 lookup must include a `reason`; do not read a full trace for every session by default.
- Open proposals and ready items explain current product understanding; they do not replace session evidence.
- Do not directly read the raw sessions store. Summarize user goals, behavior, and impact in proposals. Evidence references retain the session, relative file, and line number; do not copy large logs or specific task answers. Uncertainty is acceptable; completing a log investigation is not a submission prerequisite.

Distinguish context, observable evidence, interpretation, and candidates. Form a candidate only when evidence supports a verifiable change in user behavior and that capability is not already a stable, discoverable product contract. You may organize observations around request state, process visibility, user control, verification evidence, recovery continuity, completion handoff, and capability boundaries, but do not treat these dimensions as a predetermined checklist of problems.

A product opportunity answers "What new control, understanding, or outcome can the user obtain?", not "How should the system implement it?" For example, evidence may support "an explicit user operation receives a predictable direct response", rather than a command classifier; or "users can verify the assistant's claims about workspace state", rather than a citation protocol. Write the user capability and experience outcome first, then check that the entire output still makes sense without knowledge of internal implementation.

Do not translate internal observations directly into implementation constraints. A user need for timely steering does not require injection at the next tool boundary; controllable investigation effort does not require write-first behavior for every task with a repro. Preserve the experience goal and leave execution mechanisms to delivery.

## Proposal requirements

- Every proposal must be traceable to a session, experience reason, or user-behavior pattern.
- Titles and directions describe user-perceivable capability changes that apply across tasks.
- Submit `user_goal`, `observed_behaviors`, `experience_tension`, `product_opportunity`, `desired_outcome`, and `evidence_sessions` separately. `related_proposals` expresses context only, not a merge recommendation.
- Use `solution_level=product_capability` and `abstraction_rationale` to establish that the output stays at the level of user-perceivable product capabilities. If it prescribes trigger phrases, thresholds, protocols, state machines, or implementation steps, rewrite the product opportunity instead of marking it compliant.
- Do not provide schemas, protocols, functions, or implementation steps.
- Submit an empty list when evidence is insufficient. Do not generate candidates to meet a quantity target.

## Submission contract

After investigating, call `submit_divergent` once. `proposals` contains 0..n entries whose fields conform to the tool schema. This call creates proposals only; Owner decides whether they enter the ready queue.
\end{lstlisting}
\end{promptbox}
\begin{promptbox}{ahe\_beta/demand/divergent.py: task message}
\begin{lstlisting}
You are Product PM. The full visible user-experience index, open proposals, and ready items follow. Understand what users were trying to accomplish and the experience tensions their behavior reveals before proposing product opportunities. Existing proposals provide product-state context; overlap does not invalidate a candidate. PM goes independently to Owner; Architect does not read or absorb PM proposals. Verify using get_session_timeline with session IDs; continue to beats and steps only when needed. Do not read sessions files directly.
Submit at most 3 product insights this round, each traceable to specific session evidence.

## Experience trend (index: time, score 1-4, reason, session handle)
{experience_trend}

## Goal-aware experience summary (deterministic L1; goals, key actions, and feedback only)
{goal_summary}

## Existing proposals / ready items ({n_reference})
{reference_items}

After investigating, call `submit_divergent` once. `proposals` contains 0..n entries (an empty array if none), each separating user_goal, observed_behaviors, experience_tension, product_opportunity, desired_outcome, evidence_sessions, solution_level, abstraction_rationale, and optional related_proposals. The tool enforces the field contract.
\end{lstlisting}
\end{promptbox}

\section{Experimental Details}
\label{app:details}

\begin{table*}[t]
    \centering
    \scriptsize
    \setlength{\tabcolsep}{2pt}
    \renewcommand{\arraystretch}{1.05}
    \setlength{\abovecaptionskip}{2pt}
    \setlength{\belowcaptionskip}{2pt}
    \newcommand{\stdv}[1]{{\tiny\color{gray}$\pm$#1}}
    \caption{Full values behind Figure~\ref{fig:main}. Main comparison on 38 held-out tasks with GLM-5.3-Flash under evaluation with interaction. Each cell gives the mean, with the standard deviation in gray below it. Base Harness is the starting point of \method{}; Meta-Harness and AHE are released evolved harnesses. Pass@1: audited fraction of tasks passed with one attempt. User rating: mean rating on a 1--4 scale. Tokens: mean execution tokens in millions per task, excluding user simulation. Bold marks the best mean in each column, and ties are all bold.}
    \label{tab:main}
    \label{tab:domains}
    \resizebox{\linewidth}{!}{%
    \begin{tabular}{l @{\hspace{4pt}} rrr @{\hspace{4pt}} rrr @{\hspace{4pt}} rrr @{\hspace{4pt}} rrr @{\hspace{4pt}} rrr}
        \toprule
        & \multicolumn{3}{c}{Overall} & \multicolumn{3}{c}{Coding} & \multicolumn{3}{c}{Optimization} & \multicolumn{3}{c}{Analysis} & \multicolumn{3}{c}{Tool Use} \\
        & \multicolumn{3}{c}{\scriptsize 38 tasks} & \multicolumn{3}{c}{\scriptsize 14 tasks} & \multicolumn{3}{c}{\scriptsize 12 tasks} & \multicolumn{3}{c}{\scriptsize 7 tasks} & \multicolumn{3}{c}{\scriptsize 5 tasks} \\
        \cmidrule(lr){2-4}\cmidrule(lr){5-7}\cmidrule(lr){8-10}\cmidrule(lr){11-13}\cmidrule(lr){14-16}
        Harness
        & Pass@1\,$\uparrow$ & \begin{tabular}[b]{@{}c@{}}User\\rating\,$\uparrow$\end{tabular} & Tokens\,$\downarrow$
        & Pass@1\,$\uparrow$ & \begin{tabular}[b]{@{}c@{}}User\\rating\,$\uparrow$\end{tabular} & Tokens\,$\downarrow$
        & Pass@1\,$\uparrow$ & \begin{tabular}[b]{@{}c@{}}User\\rating\,$\uparrow$\end{tabular} & Tokens\,$\downarrow$
        & Pass@1\,$\uparrow$ & \begin{tabular}[b]{@{}c@{}}User\\rating\,$\uparrow$\end{tabular} & Tokens\,$\downarrow$
        & Pass@1\,$\uparrow$ & \begin{tabular}[b]{@{}c@{}}User\\rating\,$\uparrow$\end{tabular} & Tokens\,$\downarrow$ \\
        \midrule
        \rowcolor{gray!20} \multicolumn{16}{l}{\textbf{External harnesses}} \\
        Claude Code & 27.2\% & 2.46 & 2.89 & 38.1\% & 3.31 & 3.53 & 22.2\% & 1.53 & 3.37 & \textbf{19.0\%} & 2.05 & 1.32 & 20.0\% & 2.87 & 2.12 \\
        & \stdv{1.5} & \stdv{0.18} & \stdv{0.20} & \stdv{4.1} & \stdv{0.29} & \stdv{0.46} & \stdv{9.6} & \stdv{0.65} & \stdv{0.21} & \stdv{8.2} & \stdv{0.30} & \stdv{0.38} & \stdv{20.0} & \stdv{0.31} & \stdv{0.43} \\
        Meta-Harness & 21.1\% & 1.69 & 2.26 & 35.7\% & 2.79 & 2.86 & 19.4\% & 1.06 & 1.94 & 9.5\% & 1.20 & 2.00 & 0.0\% & 1.00 & 1.72 \\
        & \stdv{5.3} & \stdv{0.06} & \stdv{0.58} & \stdv{14.3} & \stdv{0.12} & \stdv{0.83} & \stdv{9.6} & \stdv{0.10} & \stdv{0.16} & \stdv{16.5} & \stdv{0.22} & \stdv{1.34} & \stdv{0.0} & \stdv{0.00} & \stdv{0.17} \\
        AHE & 23.7\% & 2.53 & 2.07 & 35.7\% & 3.38 & 3.55 & 16.7\% & 1.62 & \textbf{1.49} & \textbf{19.0\%} & 1.57 & 1.07 & 13.3\% & 3.60 & 0.74 \\
        & \stdv{2.6} & \stdv{0.16} & \stdv{0.18} & \stdv{0.0} & \stdv{0.21} & \stdv{0.66} & \stdv{8.3} & \stdv{0.15} & \stdv{0.43} & \stdv{8.2} & \stdv{0.52} & \stdv{0.10} & \stdv{11.5} & \stdv{0.35} & \stdv{0.10} \\
        \addlinespace
        \rowcolor{gray!20} \multicolumn{16}{l}{\textbf{Evolved from Base Harness}} \\
        Base Harness & 21.9\% & 2.59 & 2.40 & 38.1\% & 3.55 & 3.94 & 19.4\% & 1.58 & 2.01 & 9.5\% & 1.82 & 1.19 & 0.0\% & 3.35 & 0.75 \\
        & \stdv{1.5} & \stdv{0.19} & \stdv{0.25} & \stdv{10.9} & \stdv{0.21} & \stdv{0.80} & \stdv{9.6} & \stdv{0.28} & \stdv{0.44} & \stdv{16.5} & \stdv{0.37} & \stdv{0.05} & \stdv{0.0} & \stdv{0.09} & \stdv{0.35} \\
        \textbf{\method{}} & \textbf{33.3\%} & \textbf{3.04} & \textbf{1.42} & \textbf{40.5\%} & \textbf{3.81} & \textbf{1.82} & \textbf{38.9\%} & \textbf{2.31} & 1.50 & 14.3\% & \textbf{2.21} & \textbf{1.03} & \textbf{26.7\%} & \textbf{3.73} & \textbf{0.65} \\
        & \stdv{6.1} & \stdv{0.08} & \stdv{0.03} & \stdv{8.2} & \stdv{0.27} & \stdv{0.12} & \stdv{17.3} & \stdv{0.17} & \stdv{0.15} & \stdv{0.0} & \stdv{0.36} & \stdv{0.13} & \stdv{11.5} & \stdv{0.31} & \stdv{0.08} \\
        \bottomrule
    \end{tabular}%
    }
\end{table*}

\paragraph{Repeated runs.}
The main comparison runs three times and every other configuration runs once. User rating averages valid ratings only, and each run leaves out at most four sessions per harness whose user rating request failed.

\paragraph{Models and decoding.}
The execution model and the user simulator run GLM-5.3-Flash, and Triage, the Architect, and the PM run GLM-5.3. The Developers run the Claude Code CLI, which calls GLM-5.3 through an Anthropic-compatible endpoint. Evolution sets no temperature or top-p, so every role uses provider defaults. Each request of the user simulator, Base Harness, or a requirement role allows 32,768 output tokens and 256,000 input tokens, and Claude Code sets the output limit of the Developers. Evaluation runs the harness at maximum reasoning effort with thinking enabled and 65,536 output tokens per request, and the user simulator at maximum reasoning effort.

\paragraph{Rounds and sessions.}
Each evolution variant runs six rounds. Rounds 1 to 5 run sessions, both routes, admission, and delivery, and round 6 runs sessions only. A round runs one session per seed, so \usr{8} and \usrnomid{} run 8 sessions, \usr{16} runs 16, and \usr{32} runs 32, with four users at a time. These variants run 4, 4, 8, and 16 Triage instances in the same order, one Architect, and one PM, and admission calls no model. The Developers deliver four items at a time, and the next round starts only after every admitted item merges or is marked blocked.

\begin{table}[t]
    \centering
    \small
    \setlength{\tabcolsep}{4pt}
    \renewcommand{\arraystretch}{1.05}
    \setlength{\abovecaptionskip}{2pt}
    \setlength{\belowcaptionskip}{2pt}
    \caption{Issue flow of \method{} evolution by round, read from the frozen Gitea records of the run. Issues counts the issues that the 32 sessions of a round file, and Declined counts those that Triage declines, with four issues of round 5 still open. Requirements counts the requirements that Triage writes from the remaining issues together with those that ask the harness to record facts its logs miss. The Architect merges part of the requirements into consolidated requirements, and RFCs counts the changes merged into the harness, one for each consolidated requirement, PRD, and remaining requirement. Round 6 runs sessions only and is omitted.}
    \label{tab:funnel}
    \begin{tabular}{lrrrrrr}
        \toprule
        Round & Issues & Declined & Requirements & Consolidated & PRDs & RFCs \\
        \midrule
        1 & 28 & 3 & 36 & 5 & 3 & 18 \\
        2 & 26 & 6 & 23 & 5 & 3 & 18 \\
        3 & 24 & 9 & 19 & 4 & 3 & 16 \\
        4 & 23 & 6 & 18 & 5 & 2 & 10 \\
        5 & 30 & 12 & 19 & 3 & 3 & 9 \\
        \midrule
        Total & 131 & 36 & 115 & 22 & 14 & 71 \\
        \bottomrule
    \end{tabular}
\end{table}

\paragraph{Evolution cost.}
Table~\ref{tab:cost} sums recorded token components over rounds 1 to 5, and Table~\ref{tab:cost-rounds} separates the main configuration by round. For \usr{32}, the components sum to 1.92B tokens, 98\% of them input, with the Developers accounting for 68\%, about 18M per merged RFC. The harness accounts for 25\%, requirement roles for 6\%, and user interaction and review for 2\%. These shares describe the recorded components, and incomplete coverage prevents a full cost ranking across variants.

The 125 selected simulated-user sessions of \usr{32} account for 37.9M tokens across 3,367 requests, including 37.4M for interaction and 0.5M for final user ratings. One-time reviews add 3.4M. Other simulated-user requests outside the selected sessions are excluded. Requirement-role records preserve cached input and output but do not establish complete input totals. Completed development calls have usage coverage of 249/249, 368/368, 499/807, and 561/561 for \usrnomid{}, \usr{8}, \usr{16}, and \usr{32}, respectively. Another 11 failed development calls for \usrnomid{} and six for \usr{8} have no reported usage and remain uncounted.

\begin{table}[t]
    \centering
    \small
    \setlength{\tabcolsep}{4pt}
    \renewcommand{\arraystretch}{1.05}
    \setlength{\abovecaptionskip}{2pt}
    \setlength{\belowcaptionskip}{2pt}
    \caption{Recorded evolution token components in millions over rounds 1 to 5. Harness covers selected evolution sessions. User interaction and review covers simulated interaction and user ratings in selected sessions plus one-time reviews, with only reviews for \usrnomid{}. Requirement roles sums recorded cached input and output for Triage, the Architect, and the PM, including recorded retries. Developers covers selected completed development calls. The component sum is a lower bound within these scopes, and missing usage remains unknown. Totals use unrounded values.}
    \label{tab:cost}
    \begin{tabular}{lrrrrr}
        \toprule
        Variant & Harness & \begin{tabular}[b]{@{}r@{}}User interaction\\and review\end{tabular} & \begin{tabular}[b]{@{}r@{}}Requirement\\roles\end{tabular} & Developers & \begin{tabular}[b]{@{}r@{}}Component\\sum\end{tabular} \\
        \midrule
        \usrnomid{} & 232.9 & 3.6 & 25.7 & 619.1 & 881.4 \\
        \usr{8} & 113.3 & 15.8 & 36.4 & 837.2 & 1,002.7 \\
        \usr{16} & 373.6 & 46.1 & 158.3 & 1,022.1 & 1,600.1 \\
        \usr{32} & 473.1 & 41.2 & 108.7 & 1,301.7 & 1,924.8 \\
        \bottomrule
    \end{tabular}
\end{table}

\begin{table}[t]
    \centering
    \small
    \setlength{\tabcolsep}{4pt}
    \setlength{\abovecaptionskip}{2pt}
    \setlength{\belowcaptionskip}{2pt}
    \caption{Recorded token components of \method{} by round, in millions, using the scopes of Table~\ref{tab:cost}. Development coverage counts selected completed calls with usage against all selected completed calls. Calls are assigned to rounds by completion time between consecutive frozen-round boundaries. Round 6 runs sessions only and is outside this accounting window.}
    \label{tab:cost-rounds}
    \begin{tabular}{lrrrrr}
        \toprule
        Round & Harness & \begin{tabular}[b]{@{}r@{}}User interaction\\and review\end{tabular} & \begin{tabular}[b]{@{}r@{}}Requirement\\roles\end{tabular} & Developers & \begin{tabular}[b]{@{}r@{}}Development\\coverage\end{tabular} \\
        \midrule
        1 & 114.5 & 6.9 & 19.6 & 310.4 & 153/153 \\
        2 & 119.4 & 6.2 & 21.7 & 322.9 & 137/137 \\
        3 & 57.8 & 8.6 & 22.7 & 291.3 & 121/121 \\
        4 & 79.7 & 8.8 & 23.2 & 180.3 & 80/80 \\
        5 & 101.7 & 10.7 & 21.5 & 196.8 & 70/70 \\
        \midrule
        Total & 473.1 & 41.2 & 108.7 & 1,301.7 & 561/561 \\
        \bottomrule
    \end{tabular}
\end{table}

\paragraph{Evaluation budget.}
Evaluation keeps the Harbor-index time limit of each task, which the harness and the user simulator share, and runs four tasks at a time in E2B sandboxes. After the submission freezes, the user simulator rates the session as the evaluation persona from the full user-visible record within 600 seconds that do not count toward the task limit, and the user rating input excludes task scores and hidden tests.

\paragraph{Claude Code.}
Claude Code 2.1.119 from the official npm distribution runs as its native CLI on GLM-5.3-Flash through an Anthropic-compatible endpoint, with 65,536 output tokens per request. Under evaluation with interaction, each user reply resumes the same CLI session, so Claude Code keeps its conversation and its processes across replies. Its processes stop before the submission freezes, and its native session log replaces the raw API record that the other harnesses keep.

\paragraph{Meta-Harness.}
We run the harness that Meta-Harness released for Terminal-Bench 2.0, pinned at upstream revision 57fefdb. Its prompt, tool schema, model loop, and completion policy stay unchanged. An adapter only ports it to Harbor 0.24 and sends its calls to GLM-5.3-Flash through an OpenAI-compatible endpoint with the request settings of the other harnesses, and it keeps its default temperature of 0.7. Under evaluation with interaction, each user reply enters the same conversation and terminal session as a new user message, and its processes stop before the submission freezes.

\paragraph{AHE.}
We run AHE's released evolved harness \citep{lin2026ahe}, pinned at revision 182ea06, with its prescribed NexAU runtime at revision 28511d9. The adapter preserves its system prompt, shell tool, middleware, and context and iteration limits, and routes GLM-5.3-Flash calls through NexAU's native Anthropic Messages interface with the evaluation settings above. Each simulated-user reply continues the same agent instance, retaining its conversation and middleware state. We evaluate the frozen harness three times under the same tasks, persona, and time limits as the other main-comparison harnesses.

\paragraph{Ethics.}
The human A/B test releases only anonymized materials, without identity mappings, accounts, or exact submission times. All participants gave informed consent before taking part.

\subsection{Evolution mechanisms and scenario coverage}
\label{app:evolution-analysis}

\paragraph{Team-based modification turns scattered issues into cross-task changes.}
We trace every issue of \method{} in its frozen Gitea records (Table~\ref{tab:funnel}). In rounds 1 to 5, the 160 sessions file 131 issues. Triage declines 36 and rewrites 91 as requirements, and 24 more requirements ask the harness to record facts its logs miss. The Architect merges 80 of these 115 requirements into 22 consolidated requirements, and the PM writes 14 PRDs. The Developers merge the 22 consolidated requirements, the 14 PRDs, and the 35 remaining requirements as 71 RFCs. A consolidated requirement draws on sessions from a median of three evolution tasks and a PRD on sessions from a median of four, and 39 of the 71 merged changes rest on evidence from at least two tasks. No merged change names an evolution task or its repository in the runtime code, prompts, or help file of the harness, and such names appear only in RFCs, evidence files, and tests.

\begin{figure}[t]
    \centering
    \includegraphics[width=\linewidth]{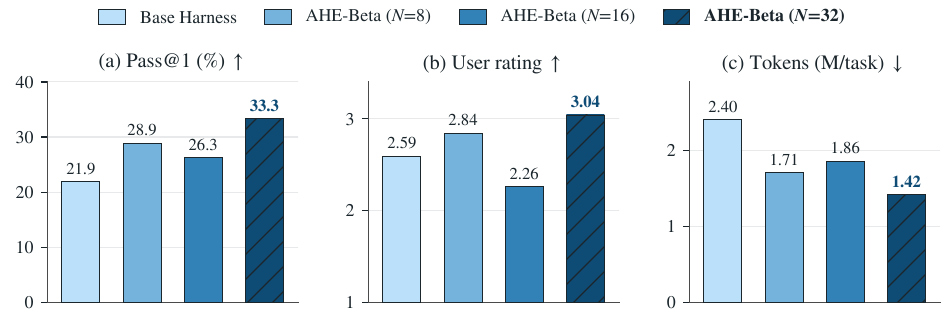}
    \caption{Scenario coverage on the 38 held-out Flash tasks. $N$ is the number of evolution seeds. Panels show (a) audited pass@1, (b) user rating from the scale floor of 1, and (c) execution tokens per task. Hatched bars mark \method{}.}
    \label{fig:coverage}
\end{figure}

\paragraph{Broader scenario coverage yields the strongest final artifact.}
\label{sec:direction}
\usr{8}, \usr{16}, and \usr{32} all improve pass@1 over Base Harness at lower token usage, and \usr{32} reaches the strongest aggregate (Figure~\ref{fig:coverage}). \usr{16} passes one task fewer than \usr{8} through a Doom timeout, and its user rating decline against Base Harness is localized to Tool Use. These independently evolved artifacts combine changes in scenario content and persona coverage, so their differences do not isolate user count. They nevertheless show useful improvements at all three coverage levels.

\paragraph{The artifacts converge on reusable support for user control.}
Source inspection finds stage checkpoints and independent verification in \usr{8}, constraint retention and scoped execution in \usr{16}, and mid-run messages and workspace-grounded progress in \usr{32}. Recorded \usr{32} stall notices followed by edits confirm that progress controls operate during evaluation. These mechanisms offer an engineering explanation of the improvements, and we do not isolate their individual effects. Appendix~\ref{app:cases} traces seven of these changes from the evolution sessions that motivated them to held-out traces.

\begin{figure}[t]
    \centering
    \includegraphics[width=\linewidth]{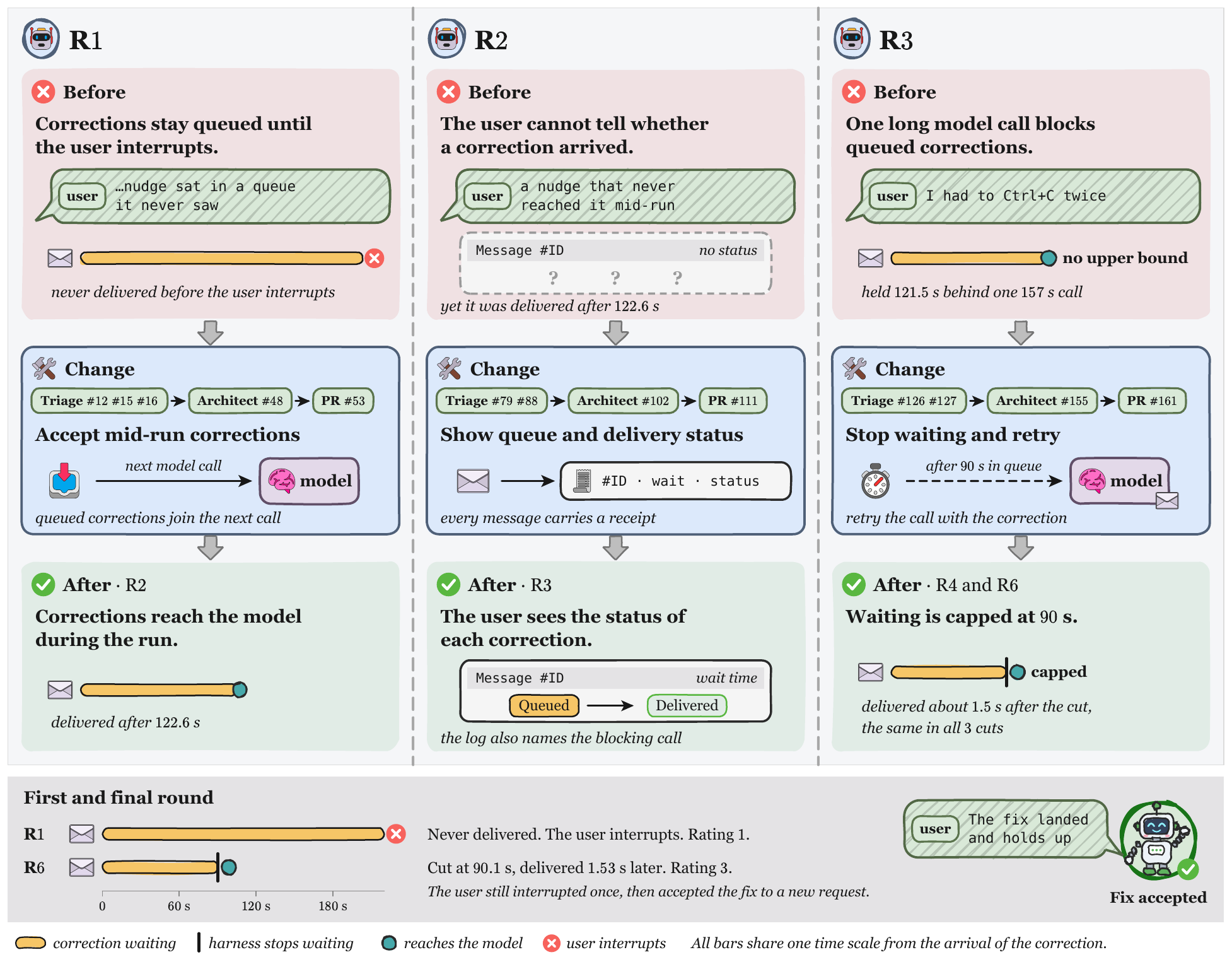}
    \caption{\textbf{From user feedback to runtime changes.} \method{}'s pydicom DS evolution case. Each column shows the problem a simulated user met in one round, the change merged after it, and the effect measured in later rounds. Bars share one time scale from the arrival of a correction. Bottom: the first and final rounds.}
    \label{fig:interaction-case}
\end{figure}

\paragraph{Case study: successive feedback improves mid-run control.}
The pydicom DS scenario shows how repeated user interaction reveals requirements left unmet by earlier fixes (Figure~\ref{fig:interaction-case}). In R1 a correction stayed queued until the user interrupted. The changes merged after R1, R2, and R3 accept corrections at model-call boundaries, expose their delivery status, and stop waiting on a long call once a correction has queued for 90\,s. Delivery took about 122\,s without a bound in R2 and R3 and about 92\,s in R4 and R6, where each correction reached the model about 1.5\,s after the cut. The user rating stayed at 1 in R4, where the user objected to the full 90\,s wait, and rose to 3 in R5 and R6. In R6 the user still interrupted once and then accepted the fix to a new request. The case connects repeated user feedback to concrete runtime changes: users need to send corrections, see their status, and have them reach the assistant while work is ongoing.

\subsection{Execution efficiency}
\label{app:execution-efficiency}

\paragraph{Token savings persist when passed tasks are held fixed.}
On tasks solved by both \method{} and its comparator, token reductions remain 53.6\% against Base Harness, 37.4\% against Meta-Harness, 41.6\% against Claude Code, and 29.7\% against AHE. Savings also remain 30.6\% against Base Harness after excluding Doom, the largest contributor. Task-matched traces show fewer model interactions, and approximately 98\% of the external baselines' recorded tokens are inputs. These diagnostics point to reduced repeated context submission as a contributor to more economical execution.

\subsection{Unmet requirements}
\label{app:unmet-requirements}

\paragraph{\method{} reaches later stages before failing.}
In several reviewed tasks, \method{} completes stages that previously blocked Base Harness but still fails a later requirement. Xarray advances from 8 of 18 to 16 of 18 passing tests, leaving two grid-boundary cases. Pandas merge progresses from a compilation failure to correct execution, then misses the speed threshold. Receipt OCR produces a structurally valid workbook where the seed produced none, but fails on a date mismatch. With DeepSeek-V4.1-Flash, Adapt 1 raises matched reference actions from 4 of 11 to 11 of 11 and fails only because of one extra email reply. Unchanged final scores therefore mask a shift from early execution failures to residual correctness, performance, extraction, and side-effect constraints.

\paragraph{The residual gap splits into unseen errors and recognized but unreachable targets.}
The remaining failures include errors that local tests miss and performance requirements that the agent states correctly but cannot satisfy. The first class needs verification evidence the harness does not yet collect, while the second needs capability the harness cannot supply by scaffolding alone.

\section{Human A/B test}
\label{app:ab}

\paragraph{Protocol.}
The study compares the frozen Base Harness and \method{}, both running GLM-5.3-Flash with the provider's default reasoning setting and an output limit of 65,536 tokens. In each attempt, a participant picks one of six tasks and works on it with two assistants labeled A and B, each with its own session and workspace. The labels hide which harness runs behind them, and A is Base Harness in 24 of the 48 valid pairs. Participants may open either side first, switch between the two, or run both in parallel, and may send follow-up requests of their own. Each side receives an execution budget of 30 minutes, or 60 minutes on T2, which participants may extend. After both sides, the participant reports the outcome of each side as completed, partial, failed, or a technical problem, rates each side from 1 to 7 on five dimensions, and states an overall preference for A, B, a tie, or neither with a written reason.
\begin{itemize}[leftmargin=*,itemsep=2pt]
\item \textbf{Usefulness.} The delivered result meets the participant's actual need.
\item \textbf{Ease of use.} The participant can easily state needs and inspect and use the result.
\item \textbf{User burden.} The task takes little attention, effort, and correction, so a higher score means less burden.
\item \textbf{Process understandability.} The participant knows what the assistant is doing and how far it has progressed.
\item \textbf{Control.} The participant can steer and correct the assistant while it proceeds on its own.
\end{itemize}

\paragraph{Tasks.}
Participants chose tasks freely, so pair counts differ across tasks (Table~\ref{tab:ab}). T1, T2, T5, and T6 build on known issues in Matplotlib, Django, and Sphinx, and their instructions ask participants not to search for an existing fix. T3 asks for a concurrency fix in a small shop service, and T4 asks for a new feature in a browser game.
\begin{itemize}[leftmargin=*,itemsep=2pt]
\item \textbf{T1 Annotation.} An arrow annotation in Matplotlib moves when outside code later mutates the NumPy array that gave its coordinates. The fix must keep the position given at creation and leave ordinary arrow annotations working.
\item \textbf{T2 Staged report.} A Django report that merges two groups of customers renders in full but fails when it exports only the points field in a given order. The participant sends five staged requests that ask the harness to diagnose the failure with run evidence and no code change, discuss a fix plan, apply a minimal fix, add descending and multi-field ordering while the harness is still running, and verify the result with regression checks. The export must keep only the selected field and leave ordinary queries unchanged.
\item \textbf{T3 Oversell.} A Python and SQLite shop service lets concurrent orders buy the last unit of stock more than once. With one unit left, concurrent requests must complete at most one order and stock must never turn negative, while ordinary orders and the existing API keep working. Optional follow-ups ask about the error on insufficient stock and about a cancellation applied twice.
\item \textbf{T4 Game feature.} In a playable browser game where a character slides through snow and avoids obstacles, the participant asks for collectible badges that disappear on contact, raise a visible counter, and reset on restart. After playing, the participant requests at least one further change of their own choice.
\item \textbf{T5 Span selector.} Enabling an interactive SpanSelector in Matplotlib expands the x axis of a curve with x between about 10 and 20 so that it includes 0. The fix must keep the axis limits stable and preserve selection and the existing horizontal and vertical selectors.
\item \textbf{T6 Literal annotation.} A strict Sphinx build treats the valid annotations \texttt{Literal[True]} and \texttt{Literal[False]} as missing class references and fails. The fix must pass the strict build and still report references that are truly missing.
\end{itemize}

\paragraph{Validity.}
Participants submitted 50 pairs, and we exclude two. In one, a side received no user message and consumed no budget after a technical problem. In the other, the tie option contradicts the written reason. The 48 valid pairs come from 25 anonymous participant labels. Completion is self-reported and receives no independent code review, so preference and ratings form the outcome of the study. In six valid pairs, three on T5, two on T6, and one on T1, the Base Harness side read the upstream patch for the task or source code containing the equivalent fix, while the audit confirms no such read on any \method{} side. This exposure favors Base Harness, yet five of these six pairs still prefer \method{}.

\begin{table}[t]
    \centering
    \small
    \setlength{\tabcolsep}{4pt}
    \renewcommand{\arraystretch}{1.05}
    \setlength{\abovecaptionskip}{2pt}
    \setlength{\belowcaptionskip}{2pt}
    \caption{Overall preference in the human A/B test by task. Pairs counts valid pairs. Without exposure removes the six valid pairs in which the Base Harness side read the upstream fix, and all submitted pairs adds back the two excluded pairs.}
    \label{tab:ab}
    \begin{tabular}{llrrrrr}
        \toprule
        Task & Codebase & Pairs & \textbf{\method{}} & Base Harness & Tie & Neither \\
        \midrule
        T1 Annotation & Matplotlib & 9 & 6 & 2 & 1 & 0 \\
        T2 Staged report & Django & 2 & 2 & 0 & 0 & 0 \\
        T3 Oversell & Shop service & 13 & 5 & 1 & 6 & 1 \\
        T4 Game feature & Browser game & 10 & 8 & 2 & 0 & 0 \\
        T5 Span selector & Matplotlib & 7 & 5 & 1 & 1 & 0 \\
        T6 Literal annotation & Sphinx & 7 & 5 & 1 & 1 & 0 \\
        \midrule
        All valid pairs & & 48 & 31 & 7 & 9 & 1 \\
        Without exposure & & 42 & 26 & 6 & 9 & 1 \\
        All submitted pairs & & 50 & 31 & 8 & 10 & 1 \\
        \bottomrule
    \end{tabular}
\end{table}

\paragraph{Results.}
Table~\ref{tab:ab-main} gives the overall summary and the five dimension scores. \method{} also receives more preferences than Base Harness on each of the six tasks (Table~\ref{tab:ab}), and excluding the six exposed pairs leaves 26 wins against 6. Ease of use and user burden already score near the top of the scale under Base Harness and gain the least. Participants report completion for 41 \method{} sides against 34 Base Harness sides, and \method{} consumes 17.8 minutes of execution budget per side on average against 23.9, where the budget also counts model waiting and timeouts.

\section{Case Studies of Evolved Changes}
\label{app:cases}

Across the six rounds the Developers merge 71 RFCs into Base Harness. Its runtime code, prompts, and tool definitions grow from about 1,100 to 18,000 lines, with the main entry file growing from 896 to 7,031 lines, 24 new runtime modules, and a system prompt growing from 18 to 63 lines. Here a harness is the program around the model that runs its tool calls, relays messages between the user and the model, and decides when a turn ends, and a turn handles one user message up to the final reply. Figures~\ref{fig:cards-success} and~\ref{fig:cards-experience} present seven of these changes as change cards, three aimed at pass@1 and four aimed at what users see and control. Each case gives the problem a simulated user met during evolution, the change, and the held-out evidence. Evolution evidence comes from the session records that each RFC cites. Held-out evidence comes from the released traces of the first main run with GLM-5.3-Flash, and run counts cover all three runs of Table~\ref{tab:main}. The traces show each change at work, and we do not isolate its effect on the outcome.

\tcbset{
    chgbase/.style={enhanced, arc=2pt, boxrule=0.6pt, left=4pt, right=4pt, top=2pt, bottom=3pt,
        toptitle=1pt, bottomtitle=1pt, fonttitle=\scriptsize\bfseries, fontupper=\scriptsize, coltitle=white},
    chgsucc/.style={chgbase, colback=gray!4, colframe=black!85},
    chgexp/.style={chgbase, colback=blue!3, colframe=blue!50!black},
}
\newcommand{\chgsec}[2][black]{\par\vspace{3pt}{\color{#1}\bfseries #2}\par\vspace{1.5pt}\hrule\vspace{2.5pt}}
\newcommand{\chgfile}[1]{\textbullet~\texttt{#1}\par}

\subsection{Changes for pass@1}

\begin{figure}[p]
\begin{tcbraster}[raster columns=2, raster equal height=rows, raster column skip=5pt, raster row skip=6pt]
\begin{tcolorbox}[chgsucc, title={Stall notice, RFCs 0009/0033/0035, level: middleware}]
\chgsec{Files}
\chgfile{transition.py}\chgfile{entry.py}\chgfile{agents/main.md}
\chgsec{What changed}
A middleware classifies every shell command as an edit or a read and tracks consecutive read-only steps and the time since the last edit. Once four read-only steps span 60\,s, or 120\,s pass without an edit, it puts an \texttt{[action-notice]} with these counts before the next model call and shows the user a \texttt{[transition]} line. If the stall lasts 60\,s past the notice, a stronger notice replaces it.
\chgsec{User problem}
The user had named the fix, yet the assistant read code for 309\,s on pydicom and about 17 minutes on sqlfluff without an edit, and the user had to kill the run or push twice.
\chgsec{Held-out evidence}
Fires on 33 of 38 tasks and escalates on 25, and an edit or new workspace content follows it 59 times on 29 tasks. Pillow filter gets the notice at 151\,s, edits 22\,s later, and passes with a 2.0x speedup at 926\,s, while Base Harness fails all three runs.
\end{tcolorbox}
\begin{tcolorbox}[chgsucc, title={Early persistence, RFC 0039, level: system prompt}]
\chgsec{Files}
\chgfile{agents/main.md}\chgfile{tools/complete\_task.tool.yaml}\chgfile{HELP.md}
\chgsec{What changed}
The prompt asks the model to write each unit that works on its own, such as a module, file, or test, to the workspace as soon as it works instead of batching all writes at the end. A supported conclusion reaches the user before checks the model added itself, which stay marked as running.
\chgsec{User problem}
On tinydb one model call streamed for 1543\,s, ran no tool, and ended in a reply of 7 characters, so no file reached the disk. On a review task the merge recommendation sat in a file for 600\,s until the user interrupted.
\chgsec{Held-out evidence}
Matrix square root writes \texttt{/app/solver.py} at 551\,s and revises it twice, and the file passes although the session ends before any reply. Base Harness keeps its code under \texttt{/tmp} and never passes.
\end{tcolorbox}
\begin{tcolorbox}[chgsucc, title={Measurement first, RFCs 0064/0066, level: runtime}]
\chgsec{Files}
\chgfile{measurement.py}\chgfile{transition.py}\chgfile{entry.py}\chgfile{agents/main.md}
\chgsec{What changed}
When the user message states a numeric target, the harness appends an ordering rule to the system prompt. The model builds the thinnest end-to-end version first, reports the first measured number and its gap to the target on a \texttt{[measured]} line, and deepens the work only after that. After 240\,s and three edits without that line, an \texttt{[action-notice]} follows.
\chgsec{User problem}
vf2-speedup-networkx requires a 5000x speedup. For 17 iterations the model built components without one measurement, and the user had to ask for the 5000x arithmetic before a C core.
\chgsec{Held-out evidence}
The target detector fires on 8 tasks, including all five AlgoTune tasks. Matrix square root measures 0.51x at $n{=}4$ and 0.94x at $n{=}512$ against a 1.01x bar at 783\,s, and both later solvers target these sizes. The \texttt{[measured]} marker itself appears on 6 tasks.
\end{tcolorbox}
\end{tcbraster}
\caption{Change cards for the three changes aimed at pass@1. Each title gives the change, the RFCs merged for it as numbered in the harness repository, and the layer it edits. Files lists the main harness files those RFCs touch. What changed describes the new behavior. User problem gives the evolution sessions that motivated the change. Held-out evidence comes from the first main run on the 38 held-out tasks with GLM-5.3-Flash.}
\label{fig:cards-success}
\end{figure}

\paragraph{Stall notice.}
The assistant sometimes kept reading code after the user had already named the change or the first step. This happened in four evolution sessions. On pydicom it read, downloaded, and tested for 309\,s without an edit until the user killed the run. On sqlfluff it read for about 17 minutes and made its first edit only after a second nudge from the user. The harness now classifies every shell command as an edit or a read. Once four consecutive read-only steps span at least 60\,s, or 120\,s pass without an edit, it injects an \texttt{[action-notice]} into the next model call. The notice states these counts and tells the model to start the named change or say what blocks it, and the user sees the same counts in a \texttt{[transition]} line. If the stall continues, a stronger notice follows. On held-out tasks the notice fires on 33 of 38 tasks and escalates on 25, and 59 notices across 29 tasks are followed directly by an edit or new workspace content. On Pillow filter, Base Harness first changes the library source at 1452\,s and reaches the 1800\,s limit without a reply. \method{} receives the notice at 151\,s and writes a benchmark script 22\,s later. After another notice at 486\,s, its next step rewrites the C loop, and at 926\,s it returns a 2.0x speedup with a passing equivalence check. This run passes, while Base Harness fails the task in all three runs.

\paragraph{Early persistence.}
Long work sometimes left nothing on disk, or its result never reached the user. On a tinydb rebuild with 11 numbered features, one model call streamed for 1543\,s without running a tool and ended with a reply of 7 characters, so no project file was written. On a code review task, the assistant wrote its merge recommendation to a file at 1189\,s but held it back for checks it had added itself, and the user interrupted at 1790\,s without receiving it. The system prompt now asks the model to write each unit that works on its own, such as a module, file, or test, to the workspace as soon as it works, and to deliver a supported conclusion before optional checks while marking those checks as still running. AlgoTune matrix square root asks for a \texttt{Solver} class in \texttt{/app/solver.py} within 1800\,s. In the first run, Base Harness keeps all its code under \texttt{/tmp} and never creates the file. \method{} writes a first solver at 551\,s, 26\,s after its first tool call, and replaces it at 1348\,s and 1680\,s with revised versions that each pass its own 66-case validation. The session reaches the time limit before any reply, yet the file on disk passes the hidden check. Across the three runs, Base Harness creates the file in only one run, at 1260\,s, and never passes, while \method{} passes both runs in which it creates the file before 1000\,s.

\paragraph{Measurement first.}
On speedup tasks the model could work for a long time without knowing how far it was from the target. On vf2-speedup-networkx the hidden check requires a 5000x geometric-mean speedup over NetworkX. For 17 iterations the model built components without measuring anything, with up to 560\,s between edits, until the user asked it to do the arithmetic for 5000x before committing to a C core. When the task text states a numeric threshold, the harness now adds an ordering rule to the first model request. The model first builds the thinnest end-to-end version, reports its first measurement and the gap to the threshold on a line marked \texttt{[measured]}, and only then deepens the work. If a turn passes 240\,s and three edits without such a line, the harness sends an \texttt{[action-notice]}. The harness detects a numeric threshold in 8 held-out tasks, including all five AlgoTune tasks. On matrix square root, the first benchmark at 783\,s measures 0.51x at $n{=}4$ and 0.94x at $n{=}512$ against a pass bar of 1.01x, and the two later solvers of the previous case target exactly these small and large sizes. The model reports these numbers in plain narration, so the \texttt{[measured]} marker itself appears on only 6 held-out tasks, none of them from AlgoTune.

\subsection{Changes for user experience}

\begin{figure}[p]
\begin{tcbraster}[raster columns=2, raster equal height=rows, raster column skip=5pt, raster row skip=6pt]
\begin{tcolorbox}[chgexp, title={Bounded correction delivery, RFCs 0004/0027/0042/0060/0068, level: middleware}]
\chgsec[blue!50!black]{Files}
\chgfile{entry.py}\chgfile{disposition.py}\chgfile{opportunity.py}\chgfile{intervention.py}
\chgsec[blue!50!black]{What changed}
A message sent during a turn joins the next model call, and each message gets a receipt with its ID, waiting time, and status. A model call that holds a message for 90\,s, or for 30\,s with no stream, is cut and retried with the message, at most twice per turn. A tool step that blocks a message for 300\,s stops the same way, and the waiting line states these bounds.
\chgsec[blue!50!black]{User problem}
On pydicom in R1 the request to make the edit now never reached the model. In R2 and R3 corrections reached it 122.6\,s and 121.5\,s after arrival, behind calls of 156\,s and 157\,s, and the user could not tell whether they had arrived.
\chgsec[blue!50!black]{Held-out evidence}
From R4 on each correction reaches the model about 1.5\,s after the cut, and the user rating rises from 1 to 3 (Figure~\ref{fig:interaction-case}). Held-out users speak only after complete replies, and human participants, who can message during a run, rate control 5.77 against 5.23.
\end{tcolorbox}
\begin{tcolorbox}[chgexp, title={Fact-anchored progress, RFCs 0002/0021/0026, level: runtime relay}]
\chgsec[blue!50!black]{Files}
\chgfile{progress.py}\chgfile{quietwatch.py}\chgfile{calibrate.py}\chgfile{entry.py}
\chgsec[blue!50!black]{What changed}
The interim text of the model now reaches the user, and every line ends with measured counts of edit commands and consecutive read-only steps. Every tool line carries a tag, namely edit, read-only, network or install, or unclassified. During silence the harness sends \texttt{[quiet]} lines with elapsed time. The prompt forbids announcing a switch to editing unless the same step edits.
\chgsec[blue!50!black]{User problem}
The model announced three times that it had gathered all facts while its read-only streak grew from 5 to 11 with no edit, and the capped status line ran out 150\,s into a turn of 1015\,s.
\chgsec[blue!50!black]{Held-out evidence}
All 589 interim lines of \method{} carry the counts, next to 764 \texttt{[quiet]} lines, while Base Harness relays no interim line. Human participants, who see these lines, rate process understandability 5.33 against 4.13.
\end{tcolorbox}
\begin{tcolorbox}[chgexp, title={Run report, RFCs 0007/0052/0053, level: runtime}]
\chgsec[blue!50!black]{Files}
\chgfile{evidence.py}\chgfile{entry.py}\chgfile{agents/main.md}
\chgsec[blue!50!black]{What changed}
Every command and its raw output go to \texttt{RUN\_REPORT.md} in the workspace as the step runs, and every turn opens with an \texttt{[evidence]} line that names the file. Turns that end with an empty reply are logged there too, and the file stays out of \texttt{git status} through the local exclude list.
\chgsec[blue!50!black]{User problem}
On kv-live-surgery only one full report reached the user, and the last request went 38 minutes without a reply. The user checked progress through file timestamps, ports, and binary comparisons, then gave up.
\chgsec[blue!50!black]{Held-out evidence}
All 41 accepted turns open with the \texttt{[evidence]} line. On matrix square root, where no reply arrives before the limit, the user can still open the record of every command that ran.
\end{tcolorbox}
\begin{tcolorbox}[chgexp, title={Readings and evidence first, RFCs 0050/0055, level: system prompt}]
\chgsec[blue!50!black]{Files}
\chgfile{agents/main.md}\chgfile{tools/complete\_task.tool.yaml}\chgfile{HELP.md}
\chgsec[blue!50!black]{What changed}
When a constraint admits materially different readings that decide whether the result meets it, the final reply states the adopted reading first, with the alternatives and their consequences. The claim hardest to recheck carries captured output or a saved artifact with a rerun command.
\chgsec[blue!50!black]{User problem}
On production-planning the readings of a 5-day plan over six dates, of due before the horizon end, and of a 24\,h freeze window lay buried in the details, so a disputed reading surfaced only after the user had audited the plan.
\chgsec[blue!50!black]{Held-out evidence}
9 of 26 first replies state an adopted reading, against none of 18 from Base Harness. On GAIA2 Adapt Hard 2 the user confirms both flagged readings and rates the session 4, citing that the assistant flagged them instead of guessing.
\end{tcolorbox}
\end{tcbraster}
\caption{Change cards for the four changes aimed at user experience. Held-out evaluation lets the user simulator speak only after complete replies, so the evidence for changes that act during a turn also comes from evolution rounds and the human A/B test (Appendix~\ref{app:ab}).}
\label{fig:cards-experience}
\end{figure}

\paragraph{Bounded correction delivery.}
A correction that the user sends while the assistant works should reach the model quickly. Figure~\ref{fig:interaction-case} follows this change through the pydicom DS seed, whose persona names a two-line fix. In R1, the assistant investigated for 309\,s without an edit, the user's message asking for the edit sat in a queue the model never read, and the user killed the run. After R1, the harness hands queued messages to the model before the next model call (RFC 0004). In R2, a correction arrived during a 156\,s model call and reached the model 122.6\,s later, while the user saw only a notice that it was queued. After R2, every message gets a receipt with its ID, its waiting time, and its delivery status (RFC 0027). In R3, a 157\,s call held a correction back for 121.5\,s, and the user interrupted twice. After R3, once a correction has waited 90\,s, or 30\,s if the model has stopped streaming, the harness cuts the model call and retries it with the correction, at most twice per turn (RFC 0042). In R4 and R6, each correction reached the model about 1.5\,s after the cut, and the user rating rose from 1 in R4 to 3 in R5 and R6. Two later RFCs apply the same bound to tool steps that hold a correction for 300\,s and to model calls that stream nothing for 120\,s, and tell a waiting user how long the wait can last and that interrupting keeps the queued message (RFCs 0060 and 0068). Held-out evaluation lets the user speak only after complete replies, so it never triggers this change. In the human A/B test, where participants may message either side at any time, control scores 5.77 for \method{} against 5.23 for Base Harness.

\paragraph{Fact-anchored progress.}
Progress messages should tell the user what actually happened, not only what the model claims. Base Harness relays only tool lines during a turn and drops the model's interim text. RFC 0002 started relaying that text, but nothing checked its claims. In one session, the model announced three times that it had gathered all the facts, while its consecutive read-only steps grew from 5 to 11 without a single edit. The status lines that carried these counts were capped at three per turn and ran out about 150\,s into a turn of about 1015\,s, leaving 19 narration lines and 29 tool results with no facts attached. The harness now attaches the edit count and consecutive read-only steps of the turn to every narration line, tags every tool line as edit, read-only, network or install, or unclassified, and emits status lines on a timer while the model is silent. The prompt also forbids announcing a switch to editing unless the same step issues an edit. All 589 narration lines that \method{} relays on held-out tasks carry these facts, while Base Harness relays no narration at all. The user simulator sees only complete replies, but the interface of the human participants shows these lines during the run, and they score process understandability 5.33 for \method{} against 4.13, the largest gain among the five dimensions.

\paragraph{Run report.}
When replies fail to arrive, the user cannot see what the assistant has run. On kv-live-surgery, the user required a report and a test on a copy at another port before any change to a live key-value server. The model respected this gate, but only one full report reached the user. The other requests ended incomplete, mostly with no visible content, and the last one went unanswered for 38 minutes. The user tracked progress through file timestamps, listening ports, and byte comparisons of binaries, then gave up and proposed a continuously updated run report. The harness now appends each command and its raw output to \texttt{RUN\_REPORT.md} in the workspace root as the command runs, and starts every turn with an \texttt{[evidence]} line that points to the file. Later RFCs also record turns that end with an empty reply in this file, and add the file to the local git exclude list after it showed up in the task's \texttt{git status} and diffs. On held-out tasks, all 41 accepted turns start with the \texttt{[evidence]} line, so even when a reply never arrives, as on matrix square root, the user can open the file and see every command that ran.

\paragraph{Readings and evidence first.}
Task text is often ambiguous, yet the harness did not say which reading it followed. On production-planning the task asked for a 5-day rolling plan from June 17 to June 22, a span of six dates. It also asked to cover jobs due before the horizon end and to dispatch nothing inside a freeze window over the first 24 hours. Each of these phrases admits two readings, and the reading decides which jobs the plan schedules. The final reply buried its readings in the details or left them out, so the user found a disputed reading only after checking the whole plan, when overturning it meant redoing the plan. Now, when a constraint admits materially different readings and the reading decides whether the deliverable meets it, the final reply opens with the adopted reading, followed by the alternatives and their consequences. For the claim that is hardest to recheck, such as passing tests, the reply attaches the captured output or names a saved artifact with a rerun command. On held-out tasks, 9 of the 26 first replies that \method{} delivers state an adopted reading, against none of the 18 from Base Harness. On GAIA2 Adapt Hard 2, the reply states that it reads a price under 1500 as strictly below 1500 and changes only the properties listed in the email. The user confirms both in the next message and rates the session 4, citing that the assistant stated its readings instead of guessing silently. Evidence first shows on Pandas merge. All three \method{} runs miss the hidden speed threshold, yet users rate each run 4, and the first-run user rating cites the equivalence check, the 5000-case differential test, and the full merge suite that the reply reports.

\end{document}